\documentclass{appolb}
\usepackage{amsmath}
\usepackage{color}
\usepackage{graphicx}
\usepackage{epsfig}

\begin{document}

\title{\hfill TPJU-02/2008\\~~ \\
Pion-to-photon transition distribution amplitudes\\
in the non-local chiral quark model}
\author{
\textbf{ Piotr Kotko{\thanks{e-mail: kotko@th.if.uj.edu.pl}
and Micha{\l} Prasza\l{}owicz{\thanks{e-mail: michal@if.uj.edu.pl}}}
 }\\
\emph{M.Smoluchowski Institute of Physics,} \\
\emph{Jagellonian University,} \\
\emph{Reymonta 4, 30-059 Krak\'{o}w, Poland.}}
\date{\today}
\maketitle

\begin{abstract}
We apply the non-local chiral quark model to study vector and axial
pion-to-photon transition amplitudes that are needed as a nonperturbative
input to estimate the cross section of pion annihilation into the real and
virtual photon. We use a simple form of the non-locality that allows to
perform all calculations in the Minkowski space and guaranties polynomiality
of the TDA's. We note only residual dependence on the precise form of the
cut-off function, however vector TDA that is symmetric in skewedness
parameter in the local quark model is no longer symmetric in the non-local
case. We calculate also the transition form-factors and compare them with
existing experimental parametrizations.
\end{abstract}

\maketitle

\section{Introduction}

\label{Intro}

Exclusive processes involving hadrons factorize in the Bjorken limit into a
hard cross section and a soft hadronic matrix element that cannot be
calculated in perturbative QCD. Those matrix elements encode nonperturbative
information on the hadronic structure. Recently Pire and Szymanowski \cite%
{PirSzym} introduced new objects of this type that describe pion-to-photon ($%
\pi2\gamma$) transition in the presence of the $q\bar{q}$ operator that in
the following will be denoted by $\Gamma$. Depending on the nature of $\Gamma
$ one can define vector or axial transition distribution amplitudes (VTDA or
ATDA respectively). In their original work Pire an Szymanowski discussed
hadron-antihadron scattering process $H\bar
{H}\rightarrow\gamma^{\ast}%
\gamma$ where the virtual photon supplies the hard scale allowing for
perturbative treatment, whereas the other photon is on mass-shell. As the
simplest case, to avoid complications with spin or multiquark bound states,
one may take pions as initial hadrons. Although experimentally difficult to
access \cite{LanPirSzy}, $\pi\pi$ scattering is of particular theoretical
interest, since pions are Goldstone bosons of broken SU(2) chiral symmetry
and their properties are to large extent determined by the symmetry
(breaking) alone rather than by the complex phenomenon of confinement.

Indeed, there exists in the literature a variety of chiral models which
involve both constituent quarks and pion degrees of freedom. In Ref.\cite%
{Tiburzi} Tiburzi used simple constituent quark model to discuss properties
of the $\pi2\gamma$ TDA's. Similar model with Pauli-Villars regularization
has been recently used by Courtoy and Noguera \cite{Courtoy} to calculate $%
\pi2\gamma$ TDA's for different ranges of kinematical variables. Finally
Ruiz Arriola and Broniowski \cite{RAB} employed the Spectral Quark Model
(SQM) for the same purpose.

One of the important ingredients of the low energy models is regularization.
Even though $\pi2\gamma$ TDA's are formally finite, regularization cannot be
simply dropped out, since it defines the scale of applicability above which
the models do not apply. In this paper we calculate $\pi2\gamma$ TDA's in
the semibosonized Nambu--Jona-Lasinio model known also as the Chiral Quark
Model ($\chi$QM) with non-local regulator. This model has been previously
used to calculate pion \cite{MPAR1}, pion and kaon \cite{Kimlocal}
distribution amplitudes (DA) and generalized parton distributions (GPD) together
with two-pion distribution amplitudes \cite{MPAR2}. Direct comparison of
local and non-local versions of the model allows to determine the influence
of the non-local regulator. In most cases rather sharp curves obtained
within the local model are smoothed down; also the endpoint behavior of
various distributions is made continuous. This phenomenon is at best
illustrated by the example of the pion distribution amplitude which in the
local model is constant over the whole support, whereas the non-local
regulator forces it to vanish in the endpoints \cite{Bochum,Bochum1}. It is
therefore of interest to investigate the role of the non-local regulator for
the $\pi2\gamma$ TDA's introduced above.

One has to remember that $\chi$QM, although devised to describe chiral
physics of Goldstone bosons, has been widely used to incorporate baryons as
chiral solitons both in local (for review see \emph{e.g.} Ref.\cite{Christov}%
) and non-local \cite{RipBroGo} cases. Generally the results of these
studies show that the soliton ceases to exist for too small constituent
quark mass $M$. The critical value of $M$ depends on the details of the
given model, however it is of the order of 300 MeV or a bit less. Typical
values of $M$ that fit well the hyperon spectrum may be as high as 420 MeV
\cite{Blotzetal}. In the present paper we adopt two distinct values of $M$:
350 MeV and 225 MeV. The latter, as we shall se in Sect.\ref{sumdis}, fits
well the slope of the $\pi 2 \gamma$ transition form-factor. It is, however,
excluded if one wants to describe baryons as chiral solitons.

Our results can be in short summarized as follows: it seems that $\pi2\gamma$
TDA's are quite robust as far as different regularization schemes (including
no cut-off at all) are concerned. On the one hand this is a welcome feature
for phenomenology, one can use them with large degree of confidence. On the
other hand they cannot be used to distinguish between different models.
However, as we shall discuss in Sect. \ref{nonlocal}, the $\xi$-symmetry of
the VTDA is no longer present in the non-local model, and the results for $%
\xi<0$ differ more from the results of the local model, than the ones for $%
\xi>0$ (where $\xi$ is the \emph{skewedness} parameter to be defined in
Sect. \ref{defkin}). We shall also see that the normalizations of VTDA and
ATDA which are equal in the local model become different in the non-local
case.

In the next section we shall discuss different chiral quark models existing
in the literature in the context of $\pi2\gamma$ TDA's. Subsequently in
Sect. \ref{NLCQM} we give a short overview of the non-local model used in
the present paper. We shall work in a symmetric kinematics that is
introduced in Sect. \ref{defkin} together with the definitions of the TDA's
in question. Calculations and results are presented in Sect. \ref{TDAs}. We
summarize and give our conclusions in Sect. \ref{sumdis}. Technical details
can be found in Appendices \ref{A}--\ref{C}.

\section{Quark models and the transition amplitudes}

\label{Models}

In order to estimate transition amplitudes one may try to construct
phenomenological ansatze that satisfy general conditions such as gauge
invariance and anomaly structure, Lorentz invariance that in our case is
equivalent to polynomiality, \emph{etc.}. For special limiting cases these
ansatze reduce to known form factors or structure functions (see \emph{e.g.}
Ref.\cite{Radyushkin:1999ms}, or in the context of $\pi2\gamma$ TDA's Ref.%
\cite{Tiburzi}). Alternatively one may try to resort to some kind of
nonperturbative calculations. Usually a good staring point is a constituent
quark model where all nonperturbative effects are parametrized in terms of a
constituent quark mass and a chirally invariant meson-quark coupling (Local
Chiral Quark Model -- L$\chi$QM). As quite useful first approximation such
model has been used in Refs.\cite{Bochum,Bochum1} to describe pion light
cone distribution amplitude or two pion distribution amplitudes \cite%
{Bochum2pi}. The results of the L$\chi$QM seem rather trivial. For example
pion light cone distribution amplitude is constant and does not vanish in
the endpoints \cite{Bochum,MPAR1}, similarly the isoscalar skewed pion
distribution amplitude is just a superposition of the $\Theta$ functions
\cite{MPAR1}. While L$\chi$QM satisfies Ward identities, it violates Lorentz
invariance by the necessary transverse UV cut-off.

Let us stress that the UV cut-off is not merely a regulator, on the
contrary, it defines the applicability domain of the model that is clearly
devised as low energy approximation to QCD. It should be therefore applied
also to the quantities that are formally finite in this limit. However, as
we shall shortly explain in some more detail, the explicit UV cut-off
results in a violation of the polynomiality which is essential for the
transition amplitudes that are discussed in this paper. Clearly a more
sophisticated regulator is needed.

One might expect that a more sophisticated regularization which preserves
Lorentz invariance would result in a more realistic shape of the pion
distribution amplitude. This is, however, not necessarily  the case. Gauge invariant
regulator is provided for example by the Spectral Quark Model (SQM) Refs.%
\cite{SQM,RAB}, where the constituent quark mass $M$ is traded for a
spectral parameter $\omega$, and all physical quantities are given in terms
of integrals over $d\omega$ with some a priori unknown spectral density $%
\rho(\omega)$. Spectral density $\rho(\omega)$ must satisfy a number of
relations that follow from the QCD Ward identities; explicit realizations of
the model with explicit form of $\rho(\omega)$ are also known \cite{SQM}.
Although at first sight theoretically attractive, SQM yields
phenomenological results that are very similar to the naive L$\chi$QM
described above.

Similarly in Ref.\cite{Davidson} a Pauli-Villars regularized
Nambu--Jona-Lasinio model was applied to calculate both pion distribution
amplitude (DA) and parton distributions in the pseudoscalar mesons. Again
pion DA at the input scale is given as a step function and does not vanish
in the endpoints. Evidently a form factor in the quark-pion vertex is needed
to \char`\"{}soften\char`\"{} the shape of the (generalized) distribution
amplitudes.

An attractive and simple way out is provided by the Non-Local Chiral Quark
Model (NL$\chi$QM) where the quark-pion coupling is given in terms of a
momentum dependent constituent quark mass $M\left( p\right) $. For small $p,$
$M(p)\rightarrow M\sim350$ MeV, whereas for $p\rightarrow\infty,$ $%
M(p)\rightarrow0$. This suppression of large momenta (remember that the
constituent quark mass $M(p)$ acts not only as a mass parameter in the
propagators, but also -- more importantly -- as a quark-meson coupling) is
enough to make the pion distribution amplitude vanish in the endpoints \cite%
{MPAR1}.

Momentum dependent constituent mass $M(p)$ preserves polynomiality, however
it violates QCD Ward identities. The latter can be easily understood by the
following example. Consider Dirac equation with a momentum dependent mass $%
M(p)$:
\begin{equation}
\left( \rlap{/}p-M(p)\right) u(p)=0   \label{Dirac}
\end{equation}
and the electromagnetic current
\begin{equation}
j^{\mu}=\overline{u}(p+q)\gamma^{\mu}u(p).   \label{curr}
\end{equation}
Naive current conservation reads
\begin{align}
q_{\mu}j^{\mu} & = \overline{u}(p+q)\left[ (\rlap{/}p+\rlap{/}q)-\rlap{/}p%
\right] u(p)  \notag \\
& = \left[ M(p+q)-M(p)\right] \,\overline{u}(p+q)u(p)\neq0.   \label{qj}
\end{align}

Modifications of the electromagnetic current $j^{\mu}$ that make the current
conserved have been proposed in 
Refs.\cite{PagelsStokar}\nocite{BallChiu,Holdom,Birse,Frank,WBnl,dor,KimNL,Noguera}
--\cite{BzdakMP}.
The discussion of the low energy theorems in the context of the instanton
model leading to the momentum dependent constituent quark mass with special
emphasis on axial anomaly can be found in Ref.\cite{Kimanomaly}. Although
these modifications, supplied by an appropriate requirements of the absence
of kinematical singularities \cite{BallChiu} fix the longitudinal part of
the pertinent non-local vertices, the transverse part remains undetermined.
When one wishes to consider electromagnetic processes, it is necessary to
assume some model for the transverse part. Therefore for the purpose of the
present work we do not consider such modifications, although we acknowledge
the fact that such a study is certainly required. Below we give an argument
in favor of such a procedure that follows from the parametrical dependence
of the current nonconservation in the instanton model of the QCD vacuum.

A non-local model leading to (\ref{Dirac}) can be \char`\"{}derived\char`\"{}
from QCD in the instanton model of the QCD vacuum \cite{DP}. In this model
the vacuum is filled with interacting instantons that stabilize in a
configuration where the mean instanton radius $\rho\sim1/(600$ MeV$)$,
whereas the typical instaton separation $R\sim1/(300$ MeV$)$. The instanton
packing fraction $(\rho/R)^{4}$ is a dynamical small parameter of the model.
The model allows to calculate $M(p)$ in Euclidean space. Setting $%
M(p)=MF^{2}(p)$ we have
\begin{equation}
F_{\text{inst.}}(p)=
2z\left[ I_{0}(z)K_{1}(z)-I_{1}(z)K_{0}(z)\right] -2I_{1}(z)K_{1}(z)
\label{Fkinst}
\end{equation}
where $z=p\rho/2$ and $M$ is the constituent mass at zero momentum.
Therefore (schematically)
\begin{eqnarray}
M(p+q)-M(p)&=&M\,\left[ F^{2}((p+q)\rho)-F^{2}(p\rho)\right]  \nonumber\\
&\simeq& M\,\frac{q\rho}{2}\frac{dF^{2}(z)}{dz}. \label{rozw}
\end{eqnarray}
Equation (\ref{rozw}) maybe viewed as an expansion in the inverse momentum $%
Q_{\text{inst.}}=2/\rho$ corresponding to the typical instanton size. Hence
for small momentum transfers (and this is certainly the domain of the
present model) the nonconservation of the vector current is parametrically
small in the inverse instanton size $Q_{\text{inst.}}$. Therefore in the
following we shall use local currents, such as (\ref{curr}), rather than the
non-local extensions, allowing for the violation of Ward identities at the
level of $q/Q_{\text{inst.}}$. The price we pay for that is wrong
normalization of the pertinent form factors, since it is fixed by the axial
anomaly. However, the dependence on the kinematical variables is almost
identical as in the models that preserve Ward identities. We shall come back
to this point in Sect.\ref{sumdis}. Finally let us stress that despite the
fact that our currents do not satisfy Ward identities the amplitudes we
calculate are gauge invariant, in the sense that they vanish when contracted
with on-shell photon momentum.

\section{Non-Local Chiral Quark Model}

\label{NLCQM}

In order to provide non-local nonperturbative regulator we employ
semibosonized Nambu--Jona-Lasinio model defined by the following action
describing quark interaction with an external meson field $U$ \cite{DP,DPrev}%
:
\begin{equation}
S_{I}=M\int\frac{d^{4}k d^{4}l}{(2\pi)^{8}}\bar{\psi}(k)F(k)U^{%
\gamma_{5}}(k-l)F(l)\psi(l)\,   \label{SI}
\end{equation}
and $U^{\gamma_{5}}(x)$ can be expanded in terms of the pion fields:
\begin{equation}
U^{\gamma_{5}}(x)=1+\frac{i}{F_{\pi}}\gamma^{5}\tau^{A}\pi^{A}(x)- \frac {1}{%
2F_{\pi}^{2}}\pi^{A}(x)\pi^{A}(x)+\ldots   \label{U}
\end{equation}
$M$ is a constituent quark mass of the order of $350$ MeV and $F(k)$ is a
momentum dependent function such that $F(0)=1$ and $F(k^{2}\rightarrow
\infty)\rightarrow0$. In what follows, for comparison, we will consider also
$M=225$ MeV

Note that (\ref{SI}) provides both momentum dependent mass of the quark
fields and the non-local quark-meson coupling. Pions act at this stage only
as auxiliary fields being -- by equations of motion -- objects composed from
quark-antiquark fields. Kinetic term for pions appears only after
integrating out the quark fields \cite{DPrev,gradient} and the proper
normalization is obtained by an appropriate choice of the cut-off function $%
F(k)$.
Here, following Refs.\cite{Bochum,MPAR1} we wish to perform all calculations
in the Minkowski space. To this end we choose: 
\begin{equation}
F(k)=\left( \frac{-\Lambda _{n}^{2}}{k^{2}-\Lambda _{n}^{2}+i\epsilon }%
\right) ^{n}  \label{Fkdef}
\end{equation}%
which reproduces reasonably well (\ref{Fkinst}) for $k^{2}<0$. Numerical
values of $\Lambda _{n}$ for different choices of $M$ are given in Table I
of Ref.\cite{MPAR1} and Table \ref{tab:Lambdas} of the present paper. Scale $%
\Lambda _{n}$ should not be confused with the typical momentum scale $Q_{0}$
(which for the original shape of $F(k)$ given by Eq.(\ref{Fkinst}) is equal
to $Q_{\text{inst}}=2/\rho$), that can be defined as the value of the momentum for
which $F(Q_{0})=$ const., say $1/2$. Then%
\begin{equation}
Q_{0}(n)=\Lambda _{n}\sqrt{\sqrt[n]{2}-1}
\end{equation}%
and does not exceed 2 GeV for the highest values of $\Lambda _{n}$.

Ansatz (\ref{Fkdef}), apart from being close to the instanton motivated
function (\ref{Fkinst}), is very practical for calculations in the Minkowski
space. Indeed, it introduces a number of complex poles in the complex
momentum plane, that can be analytically integrated over in the light cone
coordinates. Light cone coordinates are defined by two null vectors: $\tilde{%
n}=(1,0,0,1)$ and $n=(1,0,0,-1)$. In this kinematical frame any four vector $%
v$ can be decomposed as:
\begin{equation}
v^{\mu}=\frac{v^{+}}{2}\tilde{n}^{\mu}+\frac{v^{-}}{2}n^{\mu}+v_{T}^{\mu }
\label{LC}
\end{equation}
with\quad{}$v^{+}=n\cdot v,\quad v^{-}=\tilde{n}\cdot v$ and the scalar
product of two four vectors reads:
\begin{equation}
v\cdot w=\frac{1}{2}v^{+}w^{-}+\frac{1}{2}v^{-}w^{+}-\vec{v}_{T}\cdot\vec {w}%
_{T}.
\end{equation}
Therefore%
\begin{equation}
k^{2}-\Lambda_{n}^{2}=k^{-}k^{+}-\vec{k}_{T}^{\,2}-\Lambda_{n}^{2}.
\label{k2}
\end{equation}
The integration measure in the light-cone coordinates takes the following
form:%
\begin{equation}
d^{4}k=dk^{+}\, dk^{-}d^{2}\vec{k}_{T}/2 .   \label{d4k}
\end{equation}
Looking at (\ref{k2}) we see that (\ref{Fkdef}) generates a $n-$th degree
pole in the $k^{-}$ plane that can be easily integrated over. The details
can be found in Ref.\cite{MPAR1} and
in Sect.\ref{TDAs}.

\section{Definitions and kinematics}

\label{defkin}

We use the definitions of pion TDA's from \cite{Tiburzi} (our definitions
include additional $i$ phase factor)
\begin{eqnarray}
\int\frac{d\lambda}{2\pi}e^{i\lambda Xp^{+}}& \times &
\left\langle \gamma\left( P_{2},\varepsilon\right) \left\vert \overline {d}%
\left( -\frac{\lambda}{2}n\right) \gamma^{\mu}u\left( \frac{\lambda}{2}%
n\right) \right\vert \pi^{+}\left( P_{1}\right) \right\rangle \nonumber \\
& = & i\frac{i e}{2\sqrt{2}F_{\pi}p^{+}}\varepsilon^{\mu\nu\alpha\beta}%
\varepsilon_{\nu}^{\ast}p_{\alpha}q_{\beta}\, V\left( X,\xi,t\right)
\label{eq:1}
\end{eqnarray}
\begin{eqnarray}
\int\frac{d\lambda}{2\pi}e^{i\lambda Xp^{+}} & \times &
\left\langle \gamma\left( P_{2},\varepsilon\right) \left\vert \overline {d}%
\left( -\frac{\lambda}{2}n\right) \gamma^{\mu}\gamma_{5}u\left( \frac{\lambda%
}{2}n\right) \right\vert \pi^{+}\left( P_{1}\right) \right\rangle \nonumber \\
&=&i\frac{e}{2\sqrt{2}F_{\pi}p^{+}}P_{2}^{\mu}\left( q\cdot\varepsilon^{\ast
}\right) \, A\left( X,\xi,t\right) +\dots   \label{eq:2}
\end{eqnarray}
where $V\left( X,\xi,t\right) $ and $A\left( X,\xi,t\right) $ are vector and
axial TDA respectively (we use $F_{\pi}=93$ MeV). We shall use the following
\textit{symmetric} parametrization of momenta:%
\begin{align}
P_{1\mu} & =\left( 1+\xi\right) \frac{p^{+}}{2}\tilde{n}_{\mu}+\left(
1-\xi\right) \frac{p^{2}}{2p^{+}}n_{\mu}-\frac{1}{2}q_{\mu}^{T},  \notag \\
P_{2\mu} & =\left( 1-\xi\right) \frac{p^{+}}{2}\tilde{n}_{\mu}+\left(
1+\xi\right) \frac{p^{2}}{2p^{+}}n_{\mu}+\frac{1}{2}q_{\mu}^{T},
\label{P1P2}
\end{align}
where $p_{T}=0$, with
\begin{equation}
p_{\mu}=\frac{1}{2}\left( P_{1}+P_{2}\right) _{\mu}=\frac{p^{+}}{2}\tilde {n}%
_{\mu}+\frac{p^{2}}{2p^{+}}n_{\mu}
\end{equation}
Here $\xi$ denotes \emph{skewedness}. Momentum transfer 
reads:%
\begin{equation}
q_{\mu}=\left( P_{2}-P_{1}\right) _{\mu}=-\xi p^{+}\tilde{n}_{\mu}+\xi \frac{%
p^{2}}{p^{+}}n_{\mu}+q_{\mu}^{T}.   \label{q}
\end{equation}
Note that $q^{2}$ is related to $p^{2}$. Indeed, using the on mass-shell
condition (for $m_{\pi}=0$):%
\begin{equation*}
0=P_{1}^{2}=P_{2}^{2}=\left( 1-\xi^{2}\right) p^{2}-\frac{1}{4}\vec{q}%
_{T}^{\,2}=p^{2}+\frac{1}{4}q^{2}
\end{equation*}
we arrive at:
\begin{equation}
4 p^{2}=-q^{2}=-t>0.
\end{equation}
Considering the momentum transfer squared:
\begin{equation}
q^{2}=t=-4\xi^{2}p^{2}-\vec{q}_{T}^{\,2},\quad\text{or} \quad-4
\xi^{2}p^{2}=\left( t+\vec{q}_{T}^{\,2}\right) ,   \label{t}
\end{equation}
and inverting (\ref{t}), we get an important constraint:
\begin{equation}
\vec{q}_{T}^{\,2}=-(1-\xi^{2})t>0\qquad\rightarrow\qquad-1<\xi<1.
\label{qt2}
\end{equation}
One has to note, that the lower limit for $\xi$ depends on the order of
limits at $t=0$ and $m_{\pi} \ne 0$ \cite{Courtoy}. We have avoided this
ambiguity by keeping $m_{\pi}=0$ from the very beginning.

Photon polarization vector satisfies $\varepsilon^{\ast}\cdot P_{2}=0$. Dots
in Eq.(\ref{eq:2}) stand for parts which are structure independent and thus
irrelevant in our considerations \cite{Tiburzi,Courtoy}. In general TDA's do not posses any symmetry properties in $\xi$ and $X$ - in contrary to GPD's \cite{Tiburzi}.

One can define flavor diagonal VTDA's $V_{u}\left( X,\xi,t\right) $ and $%
V_{d}\left( X,\xi,t\right) $ by replacing in definition \eqref{eq:1} $\pi
^{+}$ by $\pi^{0}$ and taking operators $\overline{u}\left( -\frac{\lambda}{2%
}n\right) \gamma^{\mu}u\left( \frac{\lambda}{2}n\right) $ and $\overline {d}%
\left( -\frac{\lambda}{2}n\right) \gamma^{\mu}d\left( \frac{\lambda}{2}%
n\right) $ respectively. On the other hand matrix element $\left\langle
\gamma\left( P_{2},\varepsilon\right) \left\vert \overline{\psi}\left(
0\right) \gamma^{\mu}\psi\left( 0\right) \right\vert \pi^{0}\left(
P_{1}\right) \right\rangle $, where $\psi$ are now iso-doublets, is
parameterized by the pion-photon transition form factor controlling $%
\gamma^{\ast}\gamma\rightarrow\pi^{0}$ processes. Therefore one can derive
the sum rule \cite{PirSzym} relating this form factor to $V_{u}\left(
X,\xi,t\right) $ and $V_{d}\left( X,\xi,t\right) $. Its normalization is
fixed by axial anomaly connected with the Ward identity relating matrix
elements for transition of axial current to two photons with a similar
matrix elements of a pseudo-scalar current. This normalization together with
the conventions of Eq.(\ref{eq:1}) gives the normalization condition for the
VTDA:
\begin{equation}
{\displaystyle \int\limits_{-1}^{1}}dX\, V\left( X,\xi,t=0\right) = \frac{%
N_{c}}{2\pi^{2}}(Q_{u}+Q_{d})=\frac{1}{2\pi^{2}},   \label{normV}
\end{equation}
which is independent of $M$ and of $\xi$. The latter is related to the
polynomiality which states that the $n-$th moment of the TDA's in $X$ is a
polynomial in $\xi$ of degree not higher than $n$. Normalization (\ref{normV}%
) is automatically satisfied in the local chiral quark model which in the
chiral limit (\emph{i.e.} for $m_{\pi}=0$) gives the normalization of the
ATDA of Eq.(\ref{eq:2}) equal to the one of the VTDA:
\begin{equation}
{\displaystyle \int\limits_{-1}^{1}}dX\, A\left( X,\xi,t=0\right) = \frac{%
N_{c}}{6\pi^{2}}\left( Q_{u}-Q_{d}\right) = \frac{1}{2\pi^{2}}.
\label{normA}
\end{equation}
Note, however, that (\ref{normA}) is not fixed by the anomaly. This will be
important in the non-local model where the two normalizations are not equal
any more.

Moreover, we have the following sum rules relating vector and axial-vector
form factors with the relevant TDA's
\begin{align}
{\displaystyle \int\limits_{-1}^{1}}dX\, D\left( X,\xi,t\right) & =\frac{2%
\sqrt{2}F_{\pi}}{m_{\pi}}F_{D}\left( t\right) =2\sqrt{2}F_{\pi}F_{D}^{\chi}%
\left( t\right) ,   \label{eq:2b}
\end{align}
where $D$ stands for $V$ or $A$. In the chiral limit, which is considered in
this paper, we define $F_{V}^{\chi}\left( t\right) =F_{V}\left( t\right)
/m_{\pi}$ and similarly for $F_{A}$.

\section{Transition distribution amplitudes in the chiral quark model}

\label{TDAs}

\begin{figure}[h]
\begin{centering}
\includegraphics[scale=1.8]{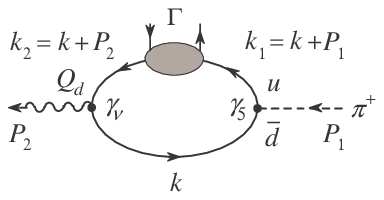}
\includegraphics[scale=1.8]{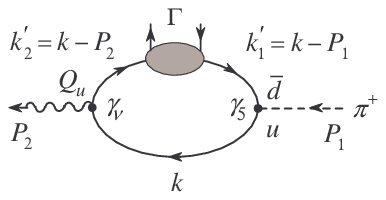}
\par\end{centering}
\caption{Feynman diagrams for $\mathcal{M}_{1}$ and $\mathcal{M}_{2}$.
Traces should be taken opposite to the momentum flow denoted by arrows. Time
flows from right to left.}
\label{fig:diagrams}
\end{figure}

Using effective action (\ref{SI}) we obtain the following expressions for
the 
matrix elements \eqref{eq:1} and \eqref{eq:2}:
\begin{eqnarray}
\int\frac{d\lambda}{2\pi}e^{i\lambda Xp^{+}}&\times&
\left\langle \gamma\left( P_{2},\varepsilon\right) \left\vert \overline {d}%
\left( -\frac{\lambda}{2}n\right) \Gamma^{\mu}u\left( \frac{\lambda}{2}%
n\right) \right\vert \pi^{+}\left( P_{1}\right) \right\rangle \nonumber \\
&=&-\frac{\sqrt{2}eMN_{c}}{F_{\pi}}\left( Q_{d}\mathcal{M}_{1}^{\mu\nu}+Q_{u}%
\mathcal{M}_{2}^{\mu\nu}\right) \varepsilon_{\nu}^{\ast},   \label{eq:3}
\end{eqnarray}
where $\Gamma^{\mu}$ is either $\gamma^{\mu}$ or $\gamma^{\mu}\gamma_{5},$ $%
N_{c}$ is number of colors, $Q_{u}$ and $Q_{d}$ are charges of quarks $u$
and $d$ respectively. Two amplitudes $\mathcal{M}_{1}$ and $\mathcal{M}_{2}$
depicted in Fig.\ref{fig:diagrams} are defined as (we omit $i\epsilon$
prescription in fermion propagators):%
\begin{eqnarray}
\mathcal{M}_{1}^{\mu\nu}&=&\int\frac{d^{4}k}{\left(2\pi\right)^{4}}\,
\delta\left(k^{+}-\left(X-1\right)p^{+}\right)F\left(k\right)F\left(k+P_{1}\right)\times  \\
& & \mathrm{Tr}\left\{ \frac{1}{\not k+\not P_{2}-M\left(k+P_{2}\right)}\Gamma^{\mu}\frac{1}{\not k+\not P_{1}-M\left(k+P_{1}\right)}\gamma_{5}\frac{1}{\not k-M\left(k\right)}\gamma^{\nu}\right\} ,\nonumber
\label{eq:4}
\end{eqnarray}
\begin{eqnarray}
\mathcal{M}_{2}^{\mu\nu}&=&\int\frac{d^{4}k}{\left(2\pi\right)^{4}}\,
\delta\left(k^{+}-\left(X+1\right)p^{+}\right)F\left(k\right)F\left(k-P_{1}\right)\times  \\
& & \mathrm{Tr}\left\{ \frac{1}{\not k-\not P_{1}-M\left(k-P_{1}\right)}\Gamma^{\mu}\frac{1}{\not k-\not P_{2}-M\left(k-P_{2}\right)}\gamma^{\nu}\frac{1}{\not k-M\left(k\right)}\gamma_{5}\right\} . \nonumber
\label{eq:5}
\end{eqnarray}

\subsection{Local case}

\label{local}

First we calculate TDA's neglecting the mass dependence upon $p$, \textit{%
i.e.} we set $M\left( p\right) =M$. It was already done within a very
similar model in Ref.~\cite{Courtoy}, where the calculations were performed
in the Minkowski space with Pauli-Villars regularization. In another
approach partially discussed in 
Refs.\cite{Bochum}\nocite{Bochum1,MPAR1}--\cite{MPAR2} one
uses the light cone coordinates (\ref{LC}) with the integration measure
given by Eq.(\ref{d4k}), performing first integration over $dk^{-}$ and then
over $dk_{T}^{2}$ with the transverse cut-off $\Lambda^{2}$. The latter is
chosen to normalize appropriately $F_{\pi}$ or alternatively to normalize
the pion distribution amplitude to 1. Introducing transverse cut-off in the
integrals defining TDA's would violate Lorentz invariance and, as a
consequence, polynomiality. This can be nicely illustrated by considering
the integral
\begin{equation}
I=\int\frac{d^{4}k}{(2\pi)^{4}}
\frac{1}{\left( k^{2}-M^{2}\right) \left( \left( k+P_{2}\right)
-M^{2}\right) \left( \left( k+P_{1}\right) -M^{2}\right) }
\end{equation}
which is related to the zeroth moment of the vector TDA and expanding it for
small $t$:

\begin{eqnarray}
I=\frac{-i}{8(2\pi)^{2}} & & \Bigg\{{\displaystyle \int}\frac{dk_{T}^{2}}{%
(k_{T}^{2}+M^{2})^{2}}+\frac{t}{4}{\displaystyle \int}dk_{T}^{2}\frac{M^{2}}{%
(k_{T}^{2}+M^{2})^{4}} + \nonumber \\
& & \frac{t}{12}\xi^{2}{\displaystyle \int}dk_{T}^{2}\frac{2k_{T}^{2}-M^{2}}{%
(k_{T}^{2}+M^{2})^{4}}+\ldots\Bigg\}.   \label{Ilocal}
\end{eqnarray}
Now, in order to make $\xi^{2}$ dependent term vanish
\begin{equation}
{\displaystyle \int\limits_{0}^{\Lambda^{2}}}dk_{T}^{2} \frac{%
2k_{T}^{2}-M^{2}}{(k_{T}^{2}+M^{2})^{4}} =-\frac{\Lambda^{2}}{%
(M^{2}+\Lambda^{2})^{3}}
\end{equation}
we have to choose $\Lambda^{2}=\infty$, what leads to:
\begin{equation}
I=\frac{-i}{8(2\pi)^{2}M^2} \Bigg\{ 1+\frac{t}{12 M^2}+\ldots \Bigg\}
\label{Ilocal2}
\end{equation}
where $\ldots$ denote higher powers of $t$.

So in order to preserve polynomiality we have to work with an infinite
transverse cut-off. In this case it is very useful to switch to
Euclidean space and use Schwinger representation for scalar propagators,
following Ref.\cite{RAB}. This allows to obtain analytical results in a very
simple way. We shall be using TDA's calculated in the local model as a
reference when discussing the results in the non-local case.

Calculating traces and combining definitions \eqref{eq:1}, \eqref{eq:2} with %
\eqref{eq:3} we get%
\begin{align}
V\left( X,\xi,t\right) & =i16M^{2}N_{c}p^{+}\left( Q_{d}\mathcal{K}_{1}+Q_{u}%
\mathcal{K}_{2}\right) ,  \label{eq:6} \\
A\left( X,\xi,t\right) & =-\frac{i}{q\cdot\varepsilon^{\ast}}%
16M^{2}N_{c}p^{+} \left( Q_{d}\mathcal{J}_{1}+Q_{u}\mathcal{J}_{2}\right) ,
\label{eq:7}
\end{align}
where%
\begin{equation}
\mathcal{K}_{1,2}=\int\frac{d^{4}k}{\left( 2\pi\right) ^{4}}
\times\frac{\delta\left( k^{+}-\left( X\mp1\right) p^{+}\right) }{\left(
\left( k\pm P_{1}\right) ^{2}-M^{2}\right) \left( \left( k\pm P_{2}\right)
^{2}-M^{2}\right) \left( k^{2}-M^{2}\right)},
\label{eq:8}
\end{equation}
\begin{equation}
\mathcal{J}_{1,2}=\int\frac{d^{4}k}{\left( 2\pi\right) ^{4}}
\times\frac{\delta\left( k^{+}-\left( X\mp1\right) p^{+}\right) \left(
\mp2k+q\right) \cdot\varepsilon^{\ast}}{\left( \left( k\pm P_{1}\right)
^{2}-M^{2}\right) \left( \left( k\pm P_{2}\right) ^{2}-M^{2}\right) \left(
k^{2}-M^{2}\right) }
\label{eq:9}
\end{equation}
with upper signs referring to subscript
\textquotedblright1\textquotedblright\ and lower signs to
\textquotedblright2\textquotedblright.

Taking the same steps as in \cite{RAB}, we get%
\begin{eqnarray}
\mathcal{K}_{1,2}&=&\frac{-i}{\left( 4\pi\right) ^{2}p^{+}}\int_{0}^{1}dy%
\int_{0}^{1-y}dz \nonumber \\
&\times&
\delta\left( y\left( 1+\xi\right) p^{+}+z\left( 1-\xi\right) p^{+}\pm\left(
X\mp1\right) \right) \frac{1}{M^{2}-yzt},
\end{eqnarray}
\begin{eqnarray}
\mathcal{J}_{1,2}&=&\frac{\pm iq\cdot\varepsilon^{\ast}}{\left( 4\pi\right)
^{2}P^{+}}\int_{0}^{1}dy\int_{0}^{1-y}dz \nonumber \\
&\times&
\delta\left( y\left( 1+\xi\right) p^{+}+z\left( 1-\xi\right) p^{+}\pm\left(
X\mp1\right) \right) \frac{1-2y}{M^{2}-yzt}.
\end{eqnarray}
While obtaining the second equation we used $n\cdot\varepsilon^{\ast}=0$ (in
the light-cone gauge), what made that expression finite. Let us notice that
from these formulae it is obvious that TDA's satisfy polynomiality
condition. Simple integration over Feynman parameters leads to the final
result, which we quote in Appendix \ref{A}.

The transition form factor we get in the local model recovers the
normalization required by the axial anomaly. Its analytical form --
calculated long time ago in Ref.\cite{Ametller} in more general kinematics
-- is given in Appendix \ref{B}.

\subsection{Non-local case}

\label{nonlocal}

In this section we take full momentum dependence of the constituent quark
mass in integrals (\ref{eq:4}) and (\ref{eq:5}). The calculations will be
done in the Minkowski space. We use method of evaluating the contour
integrals developed in Refs.\cite{MPAR1,MPAR2}.

We start from VTDA. When evaluating the trace we find not only structures
proportional to $\varepsilon^{\mu\nu\alpha\beta}\varepsilon_{\nu}^{\ast
}p_{\alpha}q_{\beta}$, which are needed, but $\varepsilon^{\mu\nu\alpha\beta
}\varepsilon_{\nu}^{\ast}n_{\alpha}P_{1\beta}$ and $\varepsilon^{\mu\nu
\alpha\beta}\varepsilon_{\nu}^{\ast}n_{\alpha}P_{2\beta}$ as well. However
they vanish when contracted with $n_{\mu}$, thus they are unphysical in the
sense that they do not give contribution to the observables. We obtain
expressions of the following form
\begin{eqnarray}
\mathcal{K}_{1,2}&=&\int\frac{d^{4}k}{\left( 2\pi\right) ^{4}}\, \frac {%
\delta\left( k^{+}-\left( X\mp1\right) p^{+}\right) F\left( k\right) F\left(
k\pm P_{1}\right) }{D\left( k\pm P_{1}\right) D\left( k\pm P_{2}\right)
D\left( k\right) } \nonumber \\
&\times& \frac{1}{M} \left\{ A_{1,2}\, M\left( k\pm P_1 \right)+B_{1,2}\,
M\left( k\pm P_2 \right)+ C_{1,2}\, M\left( k\right) \right\}
\label{int_vtda}
\end{eqnarray}
where
\begin{equation}
D\left( p\right) =p^{2}-M^{2}\left( p\right) +i\epsilon.
\end{equation}
Fuctions $A_{1,2},\, B_{1,2},\, C_{1,2}$ depend on $X,\, \xi,\, t$ and
integration variable $\vec{k}_T$. Their explicit form is given in Appendix %
\ref{C}.

Next, we have to take mass dependence on momentum given by (\ref{Fkdef}). We
choose $\Lambda_{n}$ for given $n$ in such a way that pion DA calculated in
the present model is normalized to unity. Pertinent values of $\Lambda_{n}$
are listed in Table \ref{tab:Lambdas}.

\begin{table}[h]
\begin{center}
\begin{tabular}{|c|cccc|}
\hline
$n$  & $1$  & $2$  & $3$  & $5$\\
\hline
$M=350\,\mathrm{MeV}$  & $1156$  & $1727$  & $2155$  & $2819$\\
$M=225\,\mathrm{MeV}$  & $2121$  & $3125$  & $3880$  & $5060$\\
\hline
\end{tabular}
\end{center}
\caption{Values (in MeV) of $\Lambda$ in function of $n$ for $M=350$ and
225~MeV. }
\label{tab:Lambdas}
\end{table}

Introducing dimensionless variables $\kappa=k/\Lambda,$ $\bar{P}%
_{1}=P_{1}/\Lambda,$ $\bar{P}_{2}=P_{2}/\Lambda,$ $r=M/\Lambda$ and using
\begin{align}
u_{1,2}^{\pm} & =\left( \kappa\pm\bar{P}_{1,2}\right) ^{2}-1+i\epsilon, \\
u_{3} & =\kappa^{2}-1+i\epsilon,
\end{align}
we get
\begin{eqnarray}
\mathcal{K}_{1,2}=\frac{1}{2M\Lambda^{3}}\int\frac{d^{2}\kappa_{T}d\mbox{}%
\kappa^{-}d\kappa^{+}}{\left( 2\pi\right) ^{4}} & & \frac{\delta\left(
\kappa^{+}-\left( X\mp1\right) \overline{p}^{+}\right)}{G\left(
u_{1}^{\pm}\right) G\left( u_{2}^{\pm}\right) G\left( u_{3}\right) }
\nonumber \\
& & \times \big\{ A_{1,2}\, \left( u_{1}^{\pm}\right)^{n}\left(
u_{2}^{\pm}\right)^{4n} \left( u_{3}^{\pm}\right)^{3n} \nonumber \\
& & + B_{1,2}\, \left( u_{1}^{\pm}\right)^{3n}\left( u_{2}^{\pm}\right)^{2n}
\left( u_{3}^{\pm}\right)^{3n} \nonumber \\
& &+ C_{1,2}\, \left( u_{1}^{\pm}\right)^{3n}\left( u_{2}^{\pm}\right)^{4n}
\left( u_{3}^{\pm}\right)^{n} \big\},
\label{eq:13}
\end{eqnarray}
where $G\left( u\right) =u^{4n+1}+u^{4n}-r^{2}$. Polynomial $G\left(
u\right) $ can be alternatively written in a factorized form $G\left(
u\right) =\prod_{i=1}^{4n+1}\left( u-z_{i}\right) ,$ where $z_{i}$ are roots
of equation $G\left( u\right) =0$ and can be found numerically. Note, that
if $r=0$ (\textit{i.e.} $\Lambda\rightarrow\infty$) we have $4n$ degenerate
solutions equal to zero and one equal to $-1.$ If $\Lambda$ becomes finite
the degeneracy is lifted and we have $4n+1$ solutions which in general are
complex. Integration over $d\kappa^{-}$ has to be done by the residue
theorem, thus we have to find the poles in $\kappa^{-}$ complex plane.
However, because of the imaginary part of $z_{i}\mathrm{^{\prime}s,}$ the
poles can cross the standard integration contour. This may result in
non-vanishing of the TDA's in unphysical regions. To avoid this, the
integration contour has to be modified. Detailed discussion of these
problems is given in \cite{MPAR1} and Appendix \ref{C}. After performing the
contour integrals we get:
\begin{eqnarray}
\mathcal{K}_{1,2}& = & \frac{i}{2\Lambda^{3}\bar{p}^{+}}\sum_{i,j,k=1}^{4n+1}%
\,f_{i}f_{j}f_{k}\,\int\frac{d^{2}\kappa_{T}}{\left( 2\pi\right) ^{3}}  \\
& & \frac{%
A_{1,2}z_{i}^{p_{1,2}}z_{j}^{r_{1,2}}z_{k}^{s_{1,2}}+B_{1,2}z_{i}^{a_{1,2}}z_{j}^{b_{1,2}}z_{k}^{c_{1,2}}+C_{1,2}z_{i}^{d_{1,2}}z_{j}^{e_{1,2}}z_{k}^{g_{1,2}}%
}{\left( \alpha_{1,2}z_{i}-z_{j}+\beta _{1,2}\right) \left(
\gamma_{1,2}z_{i}-z_{k}+\rho_{1,2}\right) }, \nonumber
~  \label{eq:42}
\end{eqnarray}
where
\begin{equation}
f_{i}=\prod_{j\neq i}^{4n+1}\left( z_{i}-z_{j}\right) ^{-1}
\end{equation}

\begin{figure}[h]
\begin{center}
\includegraphics[width=6cm]{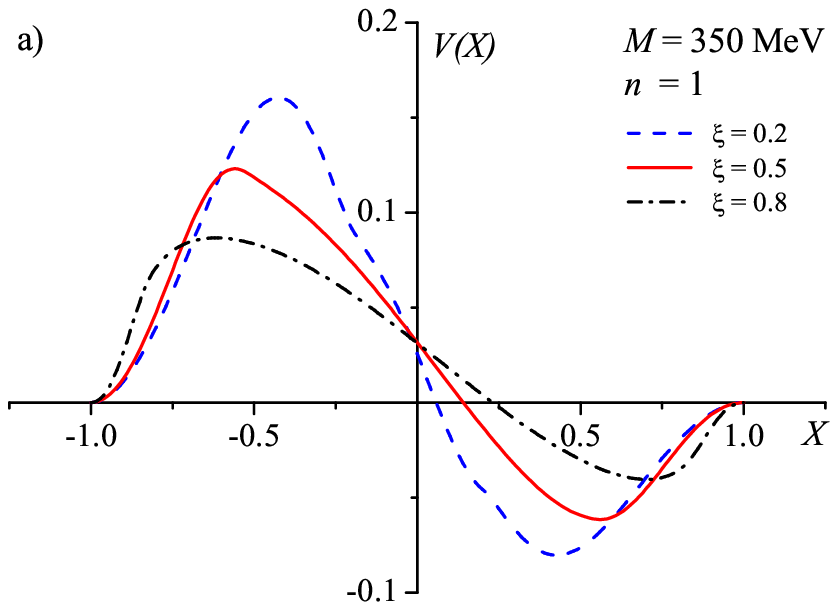}
\includegraphics[width=6cm]{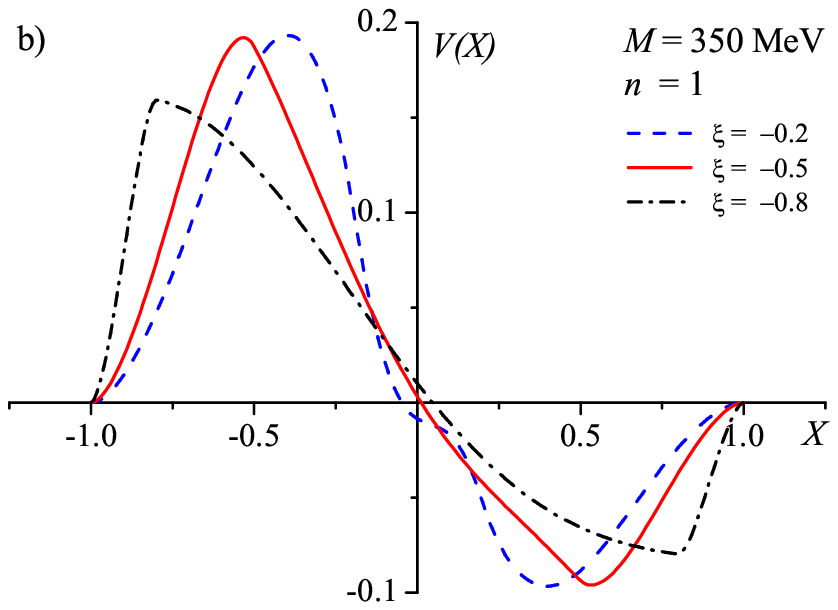}
\\
\includegraphics[width=6cm]{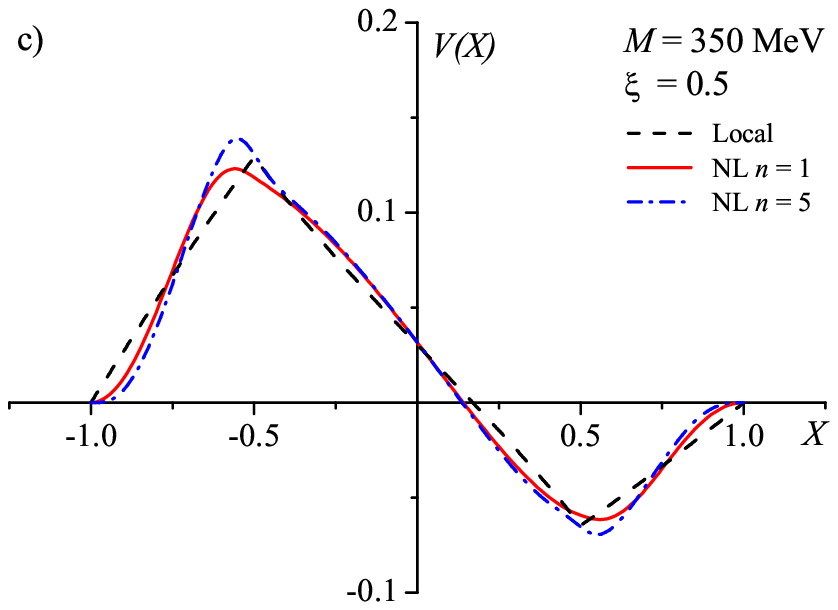}
\includegraphics[width=6cm]{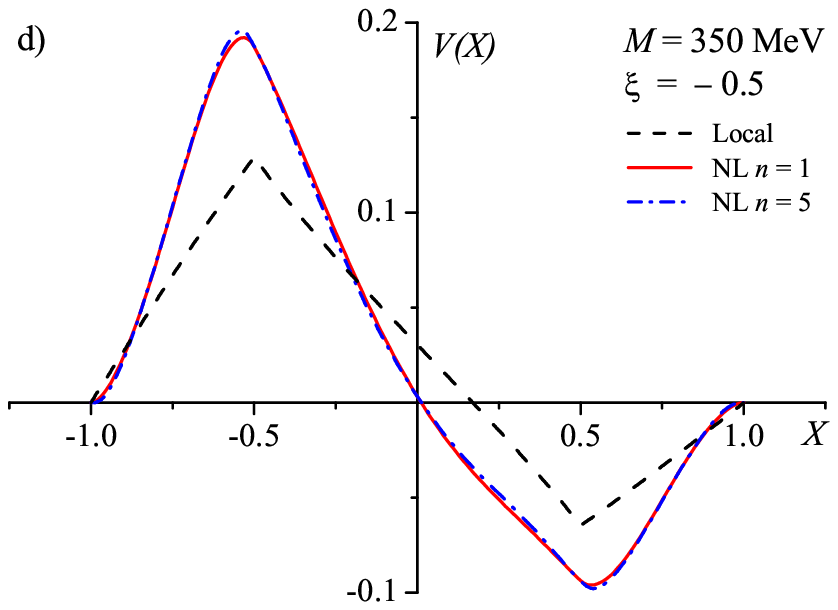}
\end{center}
\caption{Vector TDA's in the non-local model: a) for $n=1$ and $\protect\xi%
=0.2$ (dashed), $\protect\xi=0.5$ (solid), $\protect\xi=0.8$ (dash-dot); b)
for $n=1$ and $\protect\xi=-0.2$ (dashed), $\protect\xi=-0.5$ (solid), $%
\protect\xi=-0.8$ (dash-dot). Comparison of the local model: c) for $\protect%
\xi=0.5$ (dashed) with the non-local model for $n=1$ (solid) and $n=5$
(dash-dot); d) for $\protect\xi=-0.5$ (dashed) with the non-local model for $%
n=1$ (solid) and $n=5$ (dash-dot). All plots are made for constituent quark
mass $M=350\,\mathrm{MeV}$ and $t=-0.1\,\mathrm{GeV^{2}}$.}
\label{fig:VTDA-nl350}
\end{figure}

\begin{figure}[h]
\begin{center}
\includegraphics[width=6cm]{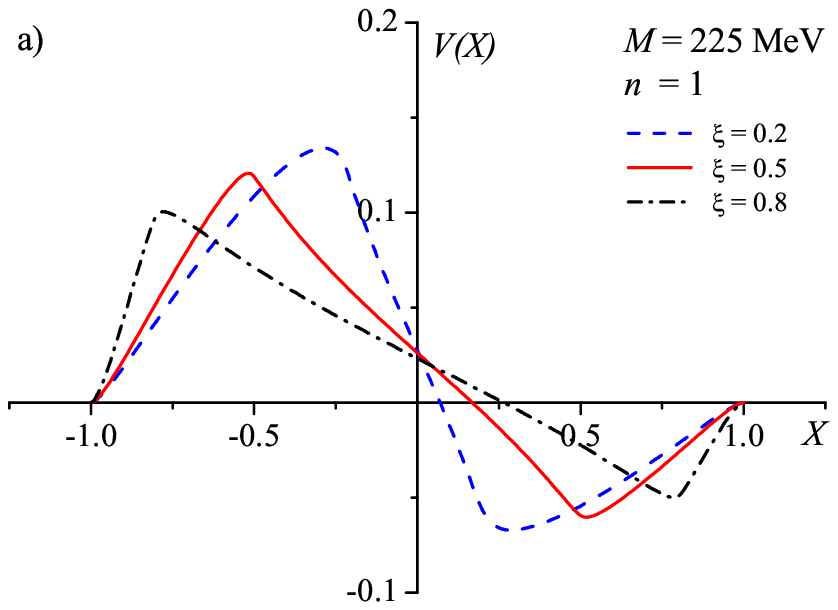}
\includegraphics[width=6cm]{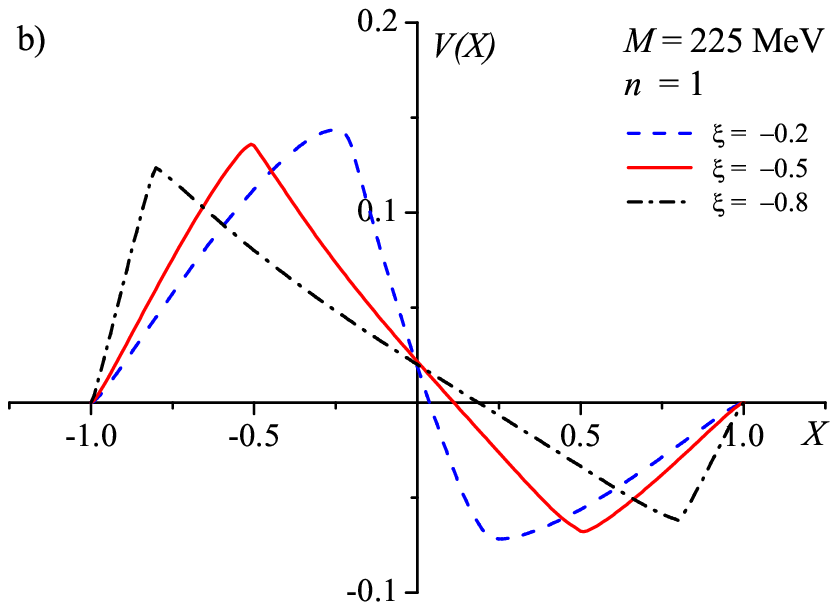}
\\
\includegraphics[width=6cm]{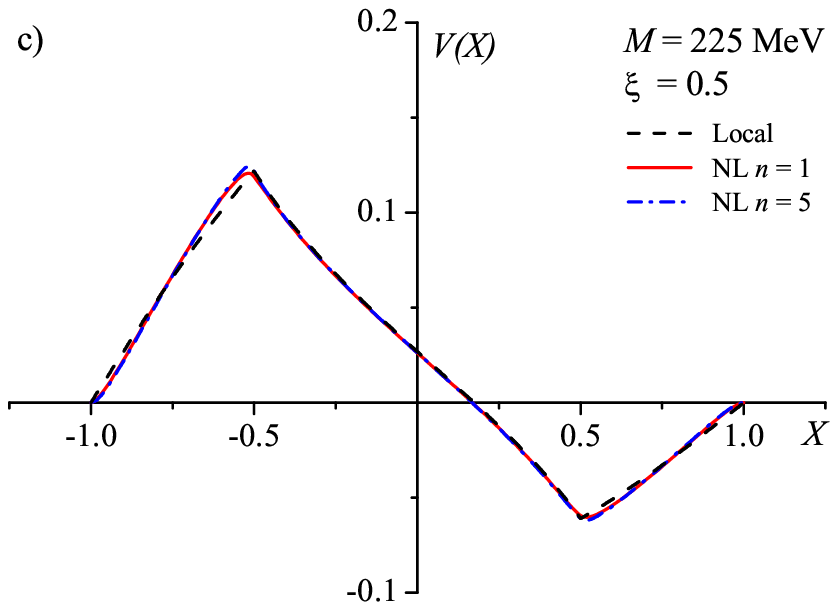}
\includegraphics[width=6cm]{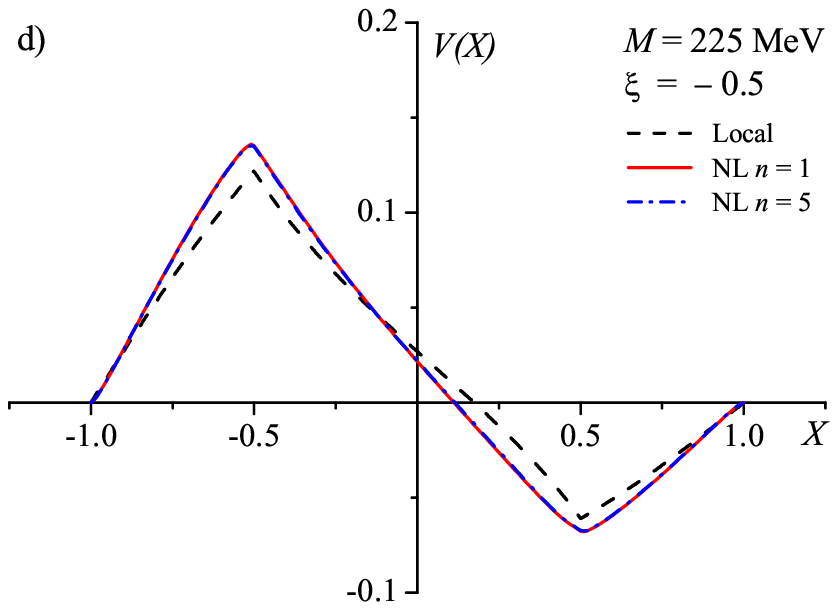}
\end{center}
\caption{Vector TDA's in the non-local model: a) for $n=1$ and $\protect\xi%
=0.2$ (dashed), $\protect\xi=0.5$ (solid), $\protect\xi=0.8$ (dash-dot); b)
for $n=1$ and $\protect\xi=-0.2$ (dashed), $\protect\xi=-0.5$ (solid), $%
\protect\xi=-0.8$ (dash-dot). Comparison of the local model: c) for $\protect%
\xi=0.5$ (dashed) with the non-local model for $n=1$ (solid) and $n=5$
(dash-dot); d) for $\protect\xi=-0.5$ (dashed) with the non-local model for $%
n=1$ (solid) and $n=5$ (dash-dot). All plots are made for constituent quark
mass $M=225\,\mathrm{MeV}$ and $t=-0.1\,\mathrm{GeV^{2}}$.}
\label{fig:VTDA-nl225}
\end{figure}

Here powers of $z_{i}$-s (denoted by Latin characters), as well as the
explicit form of the functions $\alpha_{1,2}$, $\beta_{1,2}$, $\gamma_{1,2}$%
, $\rho_{1,2}$ depend on the region of $X$ (see Appendix \ref{C}). It should
be pointed out that expressions denoted by Greek characters contain second
power of $\kappa_{T}$, while functions $A_{1,2}$, $B_{1,2}$, $C_{1,2}$ are
of first order in $\kappa_{T}$. Therefore the integration over $%
d^2\kappa_{T}=\kappa_{T}d\kappa_{T}d\theta_{T}$ is finite. Integral over $%
d\theta_{T}$ can be done analytically by integration over a unit circle,
while integral $d\kappa_{T}$ can be performed numerically.

In the non-local model we cannot recover the required normalization (\ref%
{normV}) for finite $\Lambda$. This can be understood as a consequence of
the regularization that does not respect axial anomaly. Therefore we impose
the proper normalization (\ref{normV}) by multiplying the VTDA by a suitable
correction factor $N_{V}$ as given in Table \ref{tab:corrV}.

Results for VTDA are shown in Figs.\ref{fig:VTDA-nl350} and \ref%
{fig:VTDA-nl225} for constituent masses $M=350$ and $225$ MeV respectively.
We observe that the \textquotedblright non-local\textquotedblright\ curves
are less sharp than in the local case and that their shape depends slightly
on $n$. Also the maxima are shifted from $X=\pm\xi,$ where they were placed
in local case. The non-local model has the feature that VTDA's are no more $%
\xi \text{-symmetric}$ and for $\xi<0$ it gives results that are more peaked
than in the local case and with the middle zero in a different position. The
deviation from the local model is stronger for larger constituent masses.

\begin{figure}[h]
\begin{center}
\includegraphics[width=6cm]{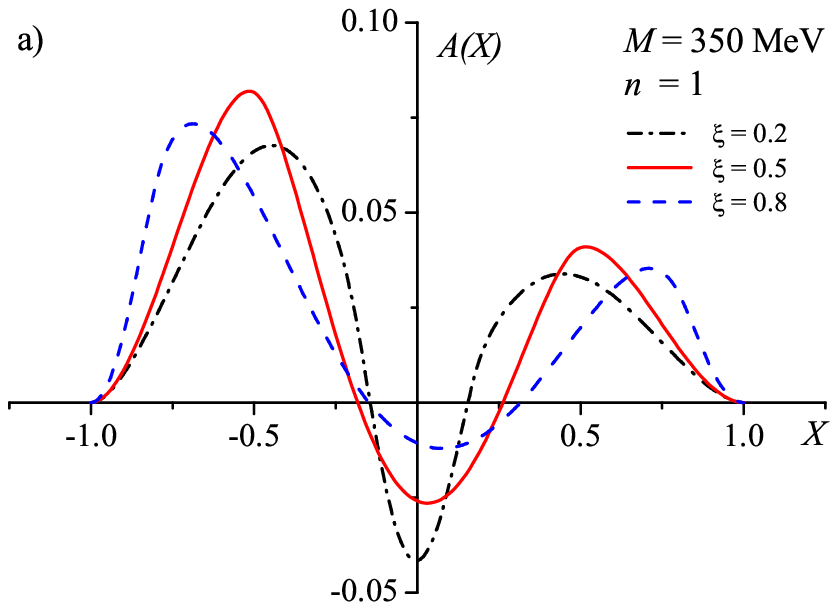}
 \includegraphics[width=6cm]{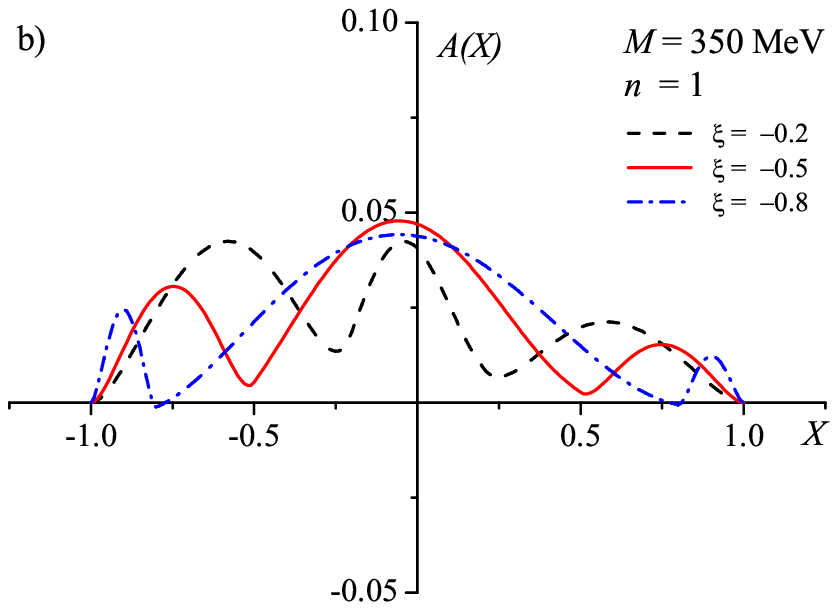} \\
\includegraphics[width=6cm]{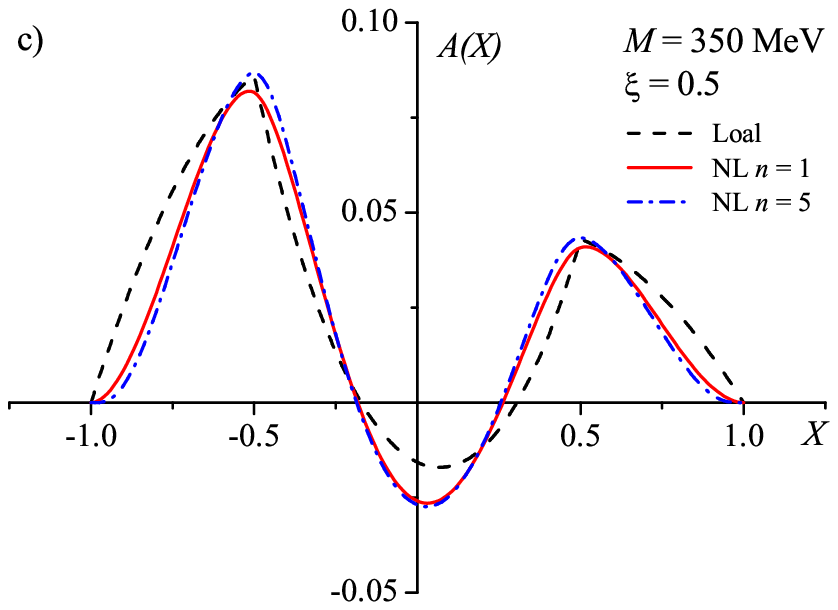}
\includegraphics[width=6cm]{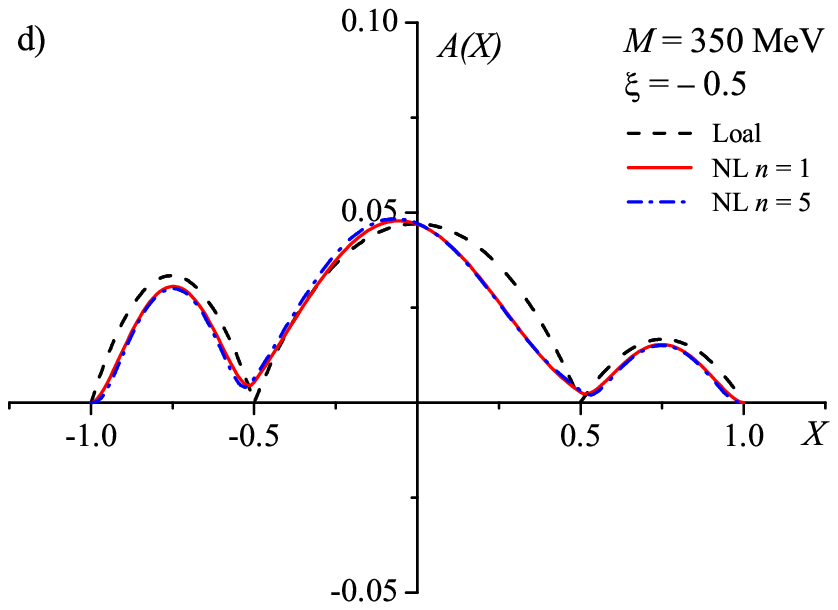}
\end{center}
\caption{Axial TDA's in the non-local model: a) for $n=1$ and $\protect\xi%
=0.2$ (dashed), $\protect\xi=0.5$ (solid), $\protect\xi=0.8$ (dash-dot); b)
for $n=1$ and $\protect\xi=-0.2$ (dashed), $\protect\xi=-0.5$ (solid), $%
\protect\xi=-0.8$ (dash-dot). Comparison of the local model: c) for $\protect%
\xi=0.5$ (dashed) with the non-local model for $n=1$ (solid) and $n=5$
(dash-dot); d) for $\protect\xi=-0.5$ (dashed) with the non-local model for $%
n=1$ (solid) and $n=5$ (dash-dot). All plots are made for constituent quark
mass $M=350\,\mathrm{MeV}$ and $t=-0.1\,\mathrm{GeV^{2}}$.}
\label{fig:ATDA-nl350}
\end{figure}

In the case of axial TDA's the algebraical steps are the same. The
complication is that after evaluating the trace we have to retain only terms
proportional to $P_{2}^{\mu}\left( q\cdot\varepsilon^{\ast }\right) $, since
all other terms are structure independent or gauge artifacts. General
expression is similar to (\ref{eq:42}), but now functions $A,\,B,\,C$
contain also second power of $\kappa_{T}.$ However, the integration over $%
d\kappa_{T}$ is finite because of the property:
\begin{equation}
\sum_{i=1}^{4n+1}z_{i}^{m}f_{i}=%
\begin{cases}
0 & \text{if}\quad m<4n \\
1 & \text{if}\quad m=4n%
\end{cases}%
\end{equation}
(see again \cite{MPAR1} and Appendix \ref{C}).

\begin{figure}[h]
\begin{center}
\includegraphics[width=6cm]{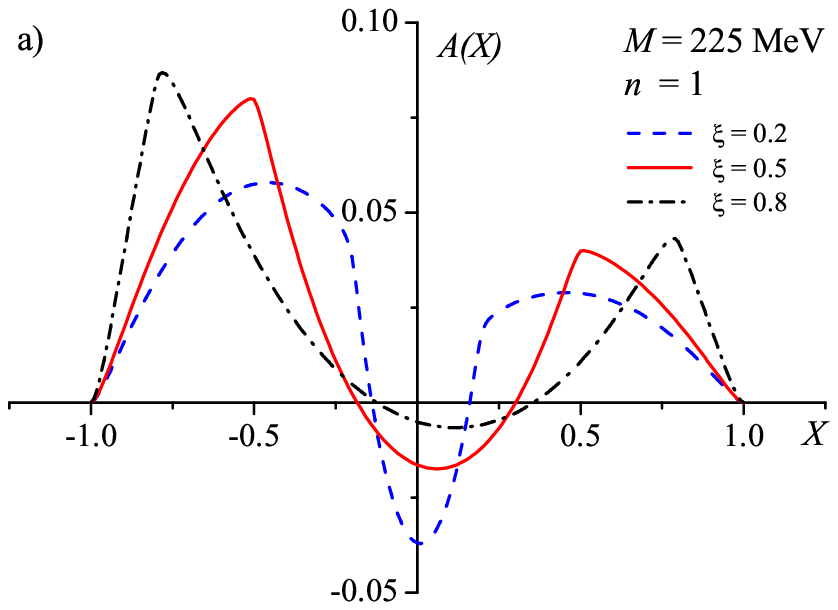}
\includegraphics[width=6cm]{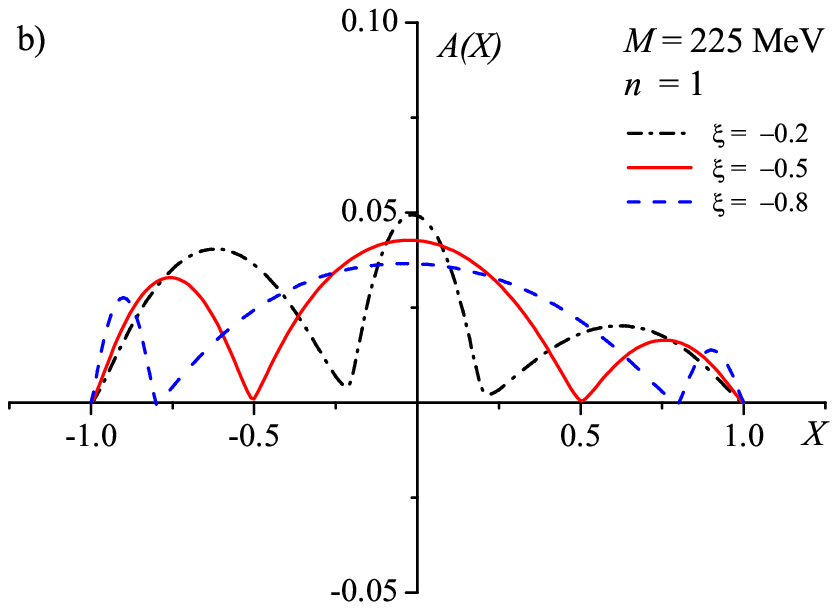}\\
\includegraphics[width=6cm]{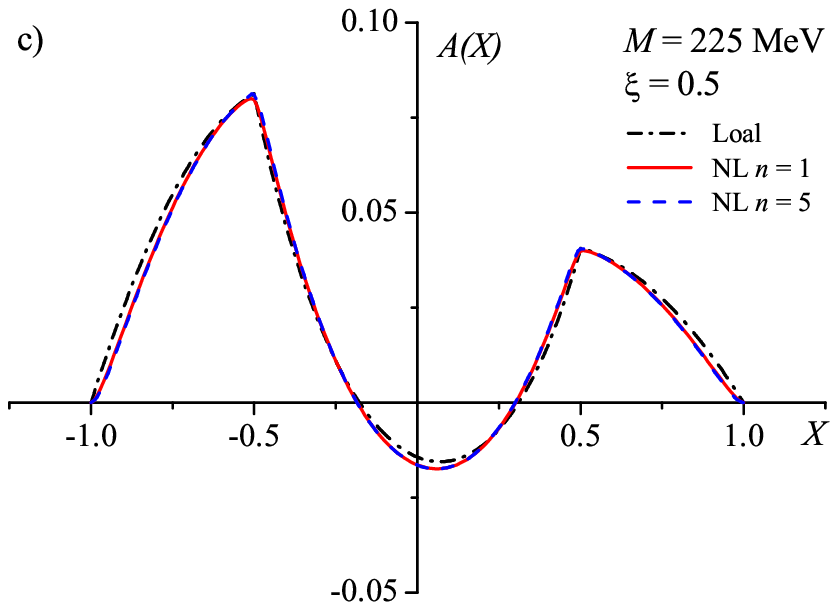}
\includegraphics[width=6cm]{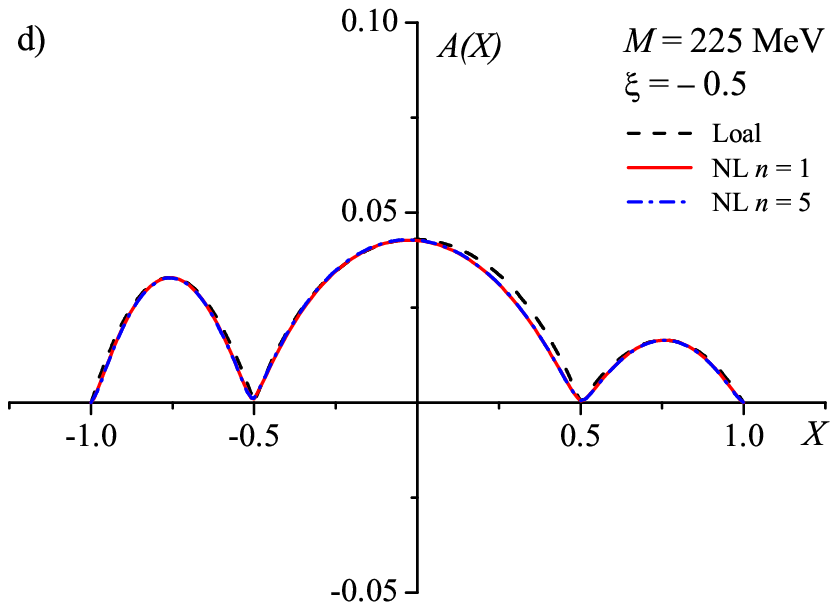}
\par\end{center}
\caption{Axial TDA's in the non-local model: a) for $n=1$ and $\protect\xi%
=0.2$ (dashed), $\protect\xi=0.5$ (solid), $\protect\xi=0.8$ (dash-dot); b)
for $n=1$ and $\protect\xi=-0.2$ (dash-dot), $\protect\xi=-0.5$ (solid), $%
\protect\xi=-0.8$ (dashed). Comparison of the local model: c) for $\protect%
\xi=0.5$ (dashed) with the non-local model for $n=1$ (solid) and $n=5$
(dash-dot); d) for $\protect\xi=-0.5$ (dashed) with the non-local model for $%
n=1$ (solid) and $n=5$ (dash-dot). All plots are made for constituent quark
mass $M=225\,\mathrm{MeV}$ and $t=-0.1\,\mathrm{GeV^{2}}$.}
\label{fig:ATDA-nl225}
\end{figure}

Numerical results for the ATDA's are shown in Figs.\ref{fig:ATDA-nl350} and %
\ref{fig:ATDA-nl225} for constituent masses $M=350$ and $225$ MeV
respectively. W see again that to a good accuracy all models (local and
fully non-local one for different $n$) give the same results both for
positive and for negative $\xi$ (note, however, small shift of the minima in
the non-local case). All curves were normalized as in the local case
according to Eq.(\ref{normA}), multiplying the calculated distribution by
correction factors listed in Table~\ref{tab:corrA}. Again the deviation from
the local model is stronger for larger constituent masses.

We have checked numerically that TDA's in the non-local model satisfy
polynomiality condition for first three moments. However, contrary to the
local case, even moments of VTDA are nonzero.

\section{Summary and discussion}

\label{sumdis}

In the present paper we have employed chiral quark model with momentum
dependent constituent quark mass to calculate pion-to-photon transition
distribution amplitudes. Before we briefly summarize our results let us
discuss the main features of the model. We have chosen momentum dependence
in the simple form given in Eq.(\ref{Fkdef}) which for Euclidean momenta
resembles $M(p)$ obtained within the instanton model of the QCD vacuum. This
form of $M(p)$ allows to perform all integrations directly in the
Minkowski space and has been previously applied to calculate pion \cite%
{MPAR1} and kaon \cite{KimNL} distribution amplitudes and two pion
distributions as well as generalized parton distribution of the pion \cite%
{MPAR2}. The main technicality that we wish to mention, consists in the
proper choice of the integration contour in the loop momentum $k^{-}$ which
is discussed in Appendix \ref{C} and can be also found in Ref.\cite{MPAR1}.

Proper choice of the integration contour guaranties that the TDA's are real
and have proper support in kinematical variables $X$ and $\xi$ defined in
Sect.\ref{defkin} and satisfy polynomiality.

Throughout this paper we have used momentum dependent mass that acts as an
UV cutoff and, at the same time, as the quark form factor within the pion.
The latter is very important for making the pion DA vanish in the endpoints
\cite{MPAR1,Bochum}. Indeed, similar calculation of the photon DA yields
distribution that is discontinuous in the endpoints \cite{Bochum1}
reflecting the point-like nature of the quark-photon coupling.

Although we have used momentum dependent mass, we have not modified currents
accordingly 
\cite{PagelsStokar}\nocite{BallChiu,Holdom,Birse,Frank,WBnl,dor,KimNL,Noguera}--\cite{BzdakMP},
and as a consequence our model violates QCD Ward identities. This violation
is however \textquotedblright mild\textquotedblright\ as it occurs at the
level $q \rho/2$ where $\rho$ is the mean instanton radius (\ref{rozw}).
Nevertheless violation of the axial Ward identity results in the wrong
normalization of the vector TDA which is fixed by the axial anomaly. In
order to get over this deficiency we have simply corrected normalization of
VTDA to the value obtained in the local model which does satisfy axial Ward
identity. The correction factors (model results should be multiplied by $%
N_{V}$ or $N_{A}$ to obtain normalization of Eqs.(\ref{normV},\ref{normA}))
are given in Tables \ref{tab:corrV} and \ref{tab:corrA} and may seem large.
Local limit can be obtained by pushing artificially $\Lambda\rightarrow\infty
$ in (\ref{Fkdef}). Obviously the same procedure applied to the pion DA
would yield pion DA constant, but with an infinite norm. In that case
correction factor would be infinite (modulo some regularization such as a
transverse cutoff for example). On this scale correction factors of the
order of 1.5 are not excessively large.

\begin{table}[h]
\begin{center}
\begin{tabular}{|c|cccc|}
\hline
$M$  & $n=1$  & $n=2$  & $n=3$  & $n=5$\\
\hline
225 MeV  & 1.151  & 1.148  & 1.147  & 1.146\\
350 MeV  & 1.487  & 1.490  & 1.490  & 1.491\\
\hline
\end{tabular}
\end{center}
\caption{Correction factors $N_{V}$ for VTDA for $M=225$ and $350$ MeV and
different values of $n$. }
\label{tab:corrV}
\end{table}

\begin{table}[h]
\begin{center}
 \begin{tabular}{|c|cccc|}
 \hline
$M$  & $n=1$  & $n=2$  & $n=3$  & $n=5$\\
\hline
225 MeV & 1.083  & 1.081  & 1.081  & 1.080\\
350 MeV  & 1.217  & 1.219  & 1.219  & 1.219\\
\hline
\end{tabular}
\end{center}
\caption{Correction factors $N_{A}$ for ATDA for $M=225$ and $350$ MeV and
different values of $n$ }
\label{tab:corrA}
\end{table}

Despite the fact that Ward identities are not satisfied our amplitudes are
gauge invariant, \emph{i.e.} they vanish when contracted with photon
momentum.

Normalization of axial TDA's is not fixed by the anomaly. However, in order
to compare them with local model we used correction factors fixing
normalization given by (\ref{normA}). This normalization overshoots
experiment by approximately a factor of 2. Such a large mismatch is common
to local quark models \cite{RAB}. The correction procedure used in this
paper to maintain normalizations (\ref{normV}) and (\ref{normA}) is to large
extent arbitrary. Taking normalizations as they come out (\emph{i.e.}
without correction factors $N_{V,A}$) would shift the axial transition form
factor at $t=0$ towards the experimental value. At the same time VTDA would
loose correct normalization. The latter, however, should be attributed to
the violation of the Ward identities in the present version of the model and
it is the violation of the axial anomaly which is responsible for the wrong
normalization in the vector case. Clearly, only a complete calculation with
the non-local currents might resolve this discrepancy.

\begin{figure}[h]
\begin{center}
\includegraphics[width=9cm]{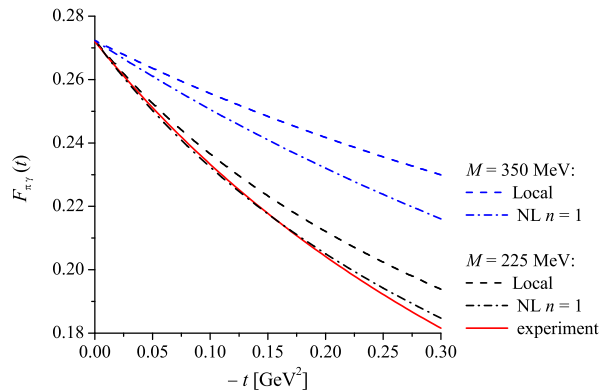}
\end{center}
\caption{Comparison of the transition form factors obtained in various
versions of the chiral quark model. Solid line represents experimental fit
of Eq.(\protect\ref{eq:a1}). Dashed lines correspond to local model,
dash-dotted lines to the non-local one with $n=1$ (there is almost no $n$
dependence). Two upper curves correspond to $M=350$ MeV, two lower ones to $%
M=225$ MeV.}
\label{fig:5}
\end{figure}

Transition form factors are defined in Eq.(\ref{eq:2b}). We show them in Fig.%
\ref{fig:5}. All calculations were performed for the constituent quark mass $%
M=350$ and $M=225\,\mathrm{MeV}$. We find that in the case of the non-local
model the transition form factor is more dumped than the one calculated in
the local version. However, for realistic constituent quark mass $M=350\,\mathrm{MeV}$
 it still falls off much slower than the
experimental curve parameterized by the function \cite{Gronberg}
\begin{equation}
F_{\pi\gamma}^{exp}\left( t\right) =\frac{F_{\pi\gamma} \left( 0\right) }{%
1-t/M_{0}^{2}},   \label{eq:a1}
\end{equation}
where $M_{0}=776\,\mathrm{MeV}$. We could get good description of
experimental data for much lower values of constituent quark mass parameter $%
M$. This can be easily understood from the approximate formula (\ref{Ilocal2}%
) where the slope reads%
\begin{equation*}
M_{0}^{2}=12M^{2}\rightarrow M=225\;\text{MeV.}
\end{equation*}
For that reason we have used $M=225$ MeV for our calculations although the
reasonable values of the constituent masses that have been used in in the
literature -- as explained in Sect. \ref{Intro} -- lie above 300 MeV.

\begin{figure}[h]
\begin{center}
\includegraphics[width=6cm]{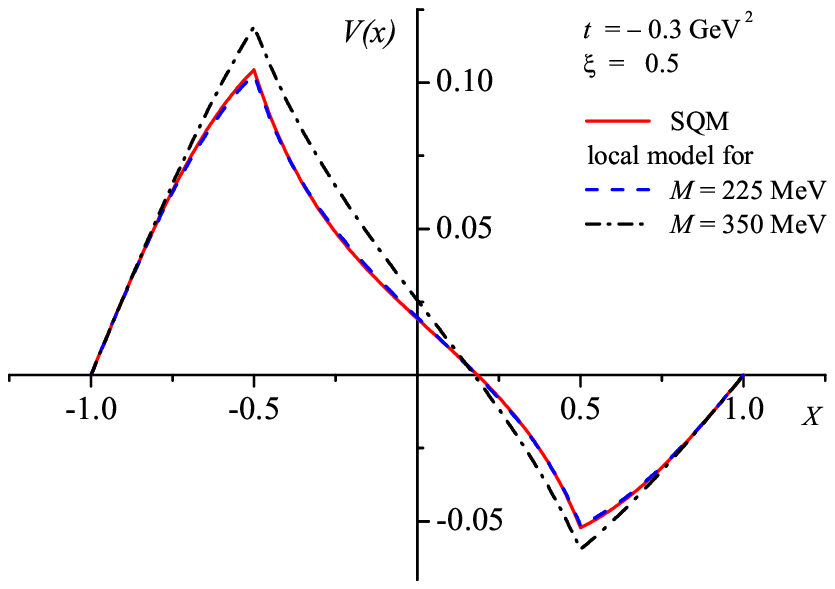}\\
\includegraphics[width=6cm]{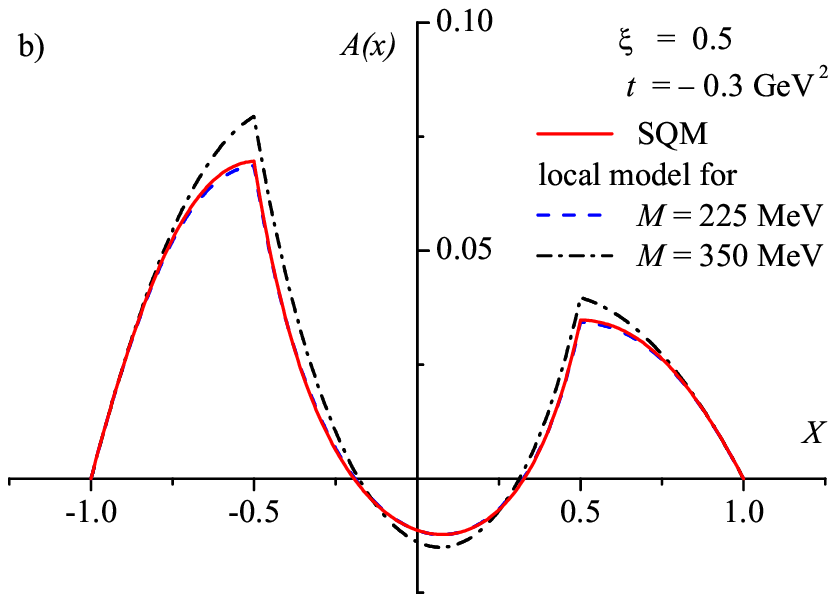}\\
\includegraphics[width=6cm]{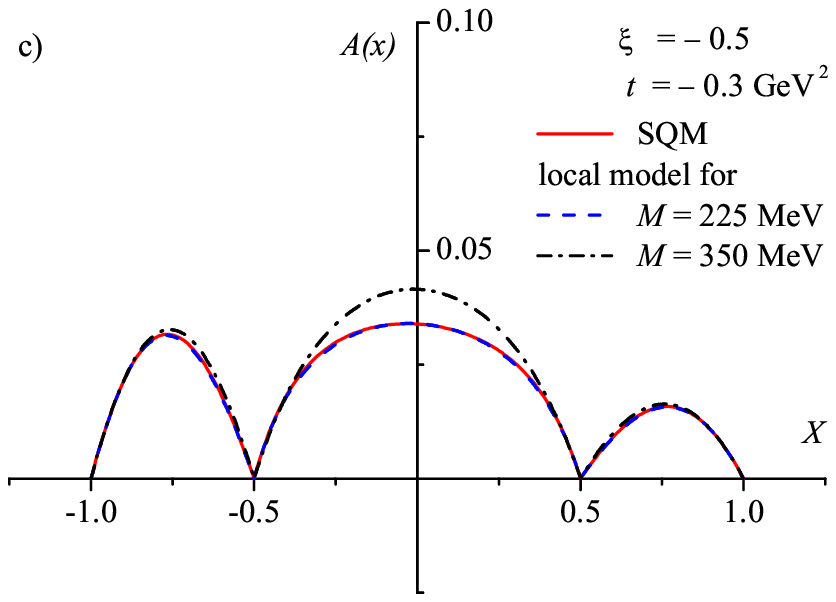}\\
\end{center}
\caption{Comparison of vector (a) and axial (b,c) TDA's obtained in the SQM
of Ref.~\protect\cite{RAB} (solid) with the local version of the present
model for $M=225$ MeV (dashed), $M= 350$ MeV (dash-dot) and $\protect\xi%
=\pm0.5$ (a), $0.5$ (b) and $-0.5$ (c). Plots are made for $t=-0.3\,\mathrm{%
MeV}$. }
\label{fig:SQM}
\end{figure}

Our calculations were performed in the symmetric kinematics defined in Sect.%
\ref{defkin} and can be directly compared with Ref.\cite{Courtoy}. There is
qualitative agreement between the local version of the present model and the
one of Ref.\cite{Courtoy} (up to an overall sign of $\xi$ for the axial case, see Ref.%
\cite{Courtoy1} ). In order to make comparison with \cite{RAB} we have
repeated their calculations in our kinematics. The results are essentially
identical to the local version of the present model, provided we take small
constituent mass. This is illustrated in Figs. \ref{fig:SQM} where we plot
vector and axial TDA's for $\xi=\pm 0.5$ and $t=-0.3$ GeV$^2$ in SQM and
local version of the present model for $M=225$ and $350$ MeV.

We see that the pion-to-photon transition amplitudes defined in Eqs.(\ref%
{eq:1}) and (\ref{eq:2}) are quite robust. Basically their shapes do not depend on
the specific model and on the regularization used. Nevertheless some small
differences between local and non-local models can be observed. The most prominent 
is the violation of the $\xi-$symmetry present in the local case for the vector TDA which can be seen in Figs.\ref%
{fig:VTDA-nl350} and \ref{fig:VTDA-nl225}. For negative $\xi$ the non-local
model gives results that are more peaked than in the local case and with the
middle zero in a different position. On the other hand ATDA's are very close
to the local case. One may be therefore confident that the shape of the
axial TDA is without doubts as shown in Figs.\ref{fig:ATDA-nl350} and \ref%
{fig:ATDA-nl225}, however normalization is not certain and should be perhaps
adjusted to the experimental value of the axial transition form factor.

From the point of view of QCD the quantities we calculate depend on a
nonperturbative scale $Q_{0}$ which, however, must not be confused neither
with the constituent mass $M$ nor with an auxiliary parameter $\Lambda$. For
$k^{2}<0$ the Ansatz (\ref{Fkdef}) should imitate $M(k)$ obtained from the
instantons. And for the latter, as explained in the Introduction, $Q_{0}
\sim2/{\rho} = 1200$~MeV. It is therefore natural to assume that $Q_{0}$ is
of the order of 1 GeV irrespectively of $M$ and $\Lambda$. The
precise definition of $Q_{0}$ is only possible within QCD and in all
effective models one can use only qualitative \emph{order of magnitude}
arguments to estimate $Q_{0}$. Discussion of this point can be found in Ref.%
\cite{MPAR1}. Once the nonperturbative scale $Q_{0}$ is fixed, our results
should be evolved to the hard scale characterizing given experimental setup
by means of the evolution equations discussed recently at length in Ref.~%
\cite{BAK}. This will be a subject of a separate study.

\vspace{0.3cm}

\noindent\textbf{Acknowledgements}

MP is grateful L. Szymanowski and to W. Broniowski and E. Ruiz Arriola for
discussions. The paper was partially supported by the Polish-German
cooperation agreement between Polish Academy of Science and DFG.

\vspace{1cm}

\newpage

\appendix

\section{Transition distribution amplitudes in the local model}

\label{A}

According to notation introduced in Sect. \ref{local} we obtained the
following expressions for VTDA (we quote results for positive $\xi$, due to
the $\xi-\text{symmetry}$)
\begin{multline}
\mathcal{K}_{1,2}=\frac{-i}{\left(4\pi\right)^{2}p^{+}}
\Bigg\{\theta\left(X+\xi\right)\theta\left(\xi-X\right)\frac{\mathrm{sign}\left(a_{1,2}-C_{1,2}\right)}{2\sqrt{AB_{1,2}}}\\
\ln\left\vert \frac{\left(a_{1,2}-C_{1,2}\right)\left(b_{1,2}+C_{1,2}\right)}{\left(a_{1,2}+C_{1,2}\right)
\left(b_{1,2}-C_{1,2}\right)}\right\vert \\
+\theta\left(1\mp X\right)\theta\left(\pm X-\xi\right)\frac{\mathrm{sign}\left(a_{1,2}-C_{1,2}\right)}{\sqrt{AB_{1,2}}}\ln\left\vert \frac{a_{1,2}-C_{1,2}}{a_{1,2}+C_{1,2}}\right\vert \Bigg\}\end{multline}
 where $A=\left( 1+\xi\right) t$, $C_{1,2}=\sqrt{B_{1,2}/A}$
and%
\begin{equation}
a_{1,2}=\frac{1\mp X}{2\left( 1+\xi\right) }
\end{equation}
\begin{equation}
b_{1,2}=\frac{\xi^{2}\mp X}{2\xi\left( 1+\xi\right) }
\end{equation}
\begin{equation}
B_{1,2}=\frac{\left( X\mp1\right) ^{2}t^{2}}{4A}-\left( 1-\xi\right) M^{2}.
\end{equation}

We remind that upper signs refer to subscript {}``1''. For ATDA
we have for positive $\xi$
\begin{multline}
\mathcal{J}_{1,2}\left(\xi>0\right)=\frac{\pm iq\cdot\varepsilon}{\left(4\pi\right)^{2}p^{+}}
\Bigg\{\theta\left(X+\xi\right)\theta\left(\xi-X\right)\frac{\left(1+D_{1,2}\right)
\mathrm{sign}\left(a_{1,2}-C_{1,2}\right)}{2\sqrt{AB_{1,2}}}\\
\times \Bigg(\ln\left|
\frac{\left(a_{1,2}-C_{1,2}\right)\left(b_{1,2}+C_{1,2}\right)}
{\left(a_{1,2}+C_{1,2}\right)\left(b_{1,2}-C_{1,2}\right)}\right|
-\frac{1}{A}\mathrm{ln}\left|\frac{Aa_{1,2}^{2}-B_{1,2}}
{Ab_{1,2}^{2}-B_{1,2}}\right|\Bigg)\\
+\theta\left(1\mp X\right)\theta\left(\mp X-\xi\right)\frac{\left(1+D_{1,2}\right)\mathrm{sign}\left(a_{1,2}-C_{1,2}\right)}
{\sqrt{AB_{1,2}}}\ln\left|\frac{a_{1,2}-C_{1,2}}{a_{1,2}+C_{1,2}}\right|\Bigg\}
\end{multline}
and for negative $\xi$
\begin{multline}
\mathcal{J}_{1,2}\left(\xi<0\right)=\frac{\pm iq\cdot\varepsilon}{\left(4\pi\right)^{2}p^{+}}
\Bigg\{\theta\left(X-\xi\right)\theta\left(-\xi-X\right)\frac{\left(1+D_{1,2}\right)\mathrm{sign}
\left(a_{1,2}-C_{1,2}\right)}{2\sqrt{AB_{1,2}}}\\
\times\Bigg(\ln\left|\frac{\left(a_{1,2}-C_{1,2}\right)
\left(b_{1,2}-C_{1,2}\right)}{\left(a_{1,2}+C_{1,2}\right)\left(b_{1,2}+C_{1,2}\right)}\right|
-\frac{1}{A}\mathrm{ln}\left|\frac{Aa_{1,2}^{2}-B_{1,2}}{Ab_{1,2}^{2}-B_{1,2}}\right|\Bigg)\\
+\theta\left(1\mp X\right)\theta\left(\pm X\mp\xi\right)\frac{\left(1+D_{1}\right)\mathrm{sign}
\left(a_{1}-C_{1}\right)}{\sqrt{AB_{1}}}
\ln\left|\frac{a_{1,2}-C_{1,2}}{a_{1,2}+C_{1,2}}\right|\Bigg\}
\end{multline}
 We introduced above 
 \begin{equation}
D_{1}=\frac{X-1}{1+\xi}\qquad D_{2}=-\frac{X+1}{1+\xi}.
\end{equation}
All the remaining notation is the same as in the VTDA case.

\section{Transition form factor in the local model}

\label{B}

We obtain the following expression for the pion-photon transition form factor%
\begin{equation}
F_{\pi\gamma}\left( t\right) =\frac{M^{2}}{2\pi^{2}F_{\pi}}\frac{1}{t}\left[
\mathrm{Li}_{2}\left( \frac{1}{\alpha_{+}}\right) +\mathrm{Li}_{2}\left(
\frac{1}{\alpha_{-}}\right) \right] ,
\end{equation}
where
\begin{equation}
\alpha_{\pm}=\frac{1}{2}\left( 1\pm\sqrt{1-\frac{4M^2}{t}}\right)
\end{equation}
and $\mathrm{Li}_{2}\left( x\right) $ is the dilogarithm function, defined
as $\mathrm{Li}_{2}\left( x\right) =-\int_{0}^{x}\,\frac{\ln\left(
1-t\right) }{t}\, dt$.

\section{TDA's in non-local model}

\label{C}

\begin{figure}[h]
\begin{center}
\includegraphics[width=7cm]{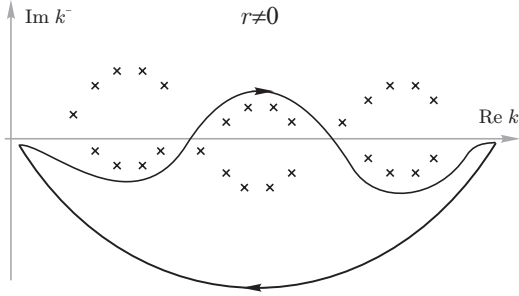}
\end{center}
\caption{Integration contour on the $\protect\kappa^{-}$ complex plane for $%
r=M/\Lambda\neq 0$. When $r\rightarrow0$ the poles from each group merge
into one pole and contour becomes standard semicircle.}
\label{contour}
\end{figure}

In Fig.\ref{contour} integration contour on the $\kappa^{-}$ complex plane
is schematically shown. As explained in more detail in main text, contour is
chosen in such a way that poles form each group cannot cross it.

In the case of VTDA we have the following general formula for the integrals (%
\ref{int_vtda})
\begin{multline}
\mathcal{K}_{1,2}=\frac{i}{2\Lambda^{3}\bar{p}^{+}}\sum_{i,j,k=1}^{4n+1}\,
f_{i}f_{j}f_{k}\,\int\frac{d^{2}\kappa_{T}}{\left( 2\pi\right) ^{3}}%
\,\epsilon_{1,2} \\
\frac{%
A_{1,2}z_{i}^{p_{1,2}}z_{j}^{r_{1,2}}z_{k}^{s_{1,2}}+B_{1,2}z_{i}^{a_{1,2}}z_{j}^{b_{1,2}}z_{k}^{c_{1,2}}+C_{1,2}z_{i}^{d_{1,2}}z_{j}^{e_{1,2}}z_{k}^{g_{1,2}}%
}{\left( \alpha_{1,2}z_{i}-z_{j}+\beta _{1,2}\right) \left(
\gamma_{1,2}z_{i}-z_{k}+\rho_{1,2}\right) }.
\end{multline}

First consider $\xi>0$ case. For $A_{1,2},\, B_{1,2},\, C_{1,2}$ we have
the following expressions:%
\begin{equation}
A_{1,2}=\mp\frac{1}{2}\left( X\mp1\right) \mp\frac{\Lambda^2\vec{\kappa}%
_{T}\cdot\vec{q}_{T}}{\left( 1+\xi\right) t}%
\end{equation}
\begin{equation}
B_{1,2}=\pm\frac{1}{2}\left( X\mp1\right) \mp\frac{\Lambda^2\vec{\kappa}%
_{T}\cdot\vec{q}_{T}}{\left( 1-\xi\right) t}%
\end{equation}
\begin{equation}
C_{1,2}=1\pm\frac{2\Lambda^2\vec{\kappa}_{T}\cdot\vec{q}_{T}}{\left(
1-\xi^{2}\right) t}.
\end{equation}

The explicit form of $\alpha_{1,2},\,%
\beta_{1,2},\,\gamma_{1,2},\,\rho_{1,2}$ and $\epsilon_{1,2}$ depends on the
region of the support under consideration. We introduce%
\begin{equation}
u_{a}^{1,2}=\pm\frac{\left(X\mp1\right)\left(1-\xi\right)t}{4\Lambda^{2}\bar{%
p}^{+} \left(X\pm\xi\right)}+\frac{\kappa_{T}^{2}\mp\vec{\kappa}_{T}\cdot%
\vec{\bar{q}}_{T}+1} {\bar{p}^{+}\left(X\pm\xi\right)}
\end{equation}
\begin{equation}
u_{b}^{1,2}=\pm\frac{\left(X\mp1\right)\left(1+\xi\right)t}{4\Lambda^{2}\bar{%
p}^{+} \left(X\mp\xi\right)}+\frac{\kappa_{T}^{2}\pm\vec{\kappa}_{T}\cdot%
\vec{\bar{q}}_{T}+1} {\bar{p}^{+}\left(X\mp\xi\right)}
\end{equation}
\begin{equation}
u_{c}^{1,2}=\frac{\kappa_{T}^{2}+1}{\bar{p}^{+}\left(X\mp1\right)}.
\end{equation}

In every region we have:

\begin{itemize}
\item $-1\leq X<-\xi$

\begin{equation}
\epsilon_{1}=0,\quad\epsilon_{2}=-\frac{1}{X+1}  \label{app1}
\end{equation}
\begin{equation}
\alpha_{2}=\frac{X+\xi}{X+1},\quad\gamma_{2}=\frac{X-\xi}{X+1}
\end{equation}
\begin{equation}
\beta_{2}=\bar{p}^{+}\left(X+\xi\right)\left(u_{c}^{2}-u_{b}^{2}\right)
\end{equation}
\begin{equation}
\rho_{2}=\bar{p}^{+}\left(X-\xi\right)\left(u_{c}^{2}-u_{a}^{2}\right)
\end{equation}

\item $-\xi\leq X<\xi$

\begin{equation}
\epsilon_{1,2}=\mp\frac{1}{X\pm\xi}  \label{app2}
\end{equation}
\begin{equation}
\alpha_{1,2}=\frac{X\mp\xi}{X\pm\xi},\quad\gamma_{1,2}=\frac{X\mp1}{X\pm\xi}
\label{app3}
\end{equation}

\begin{equation}
\beta_{1,2}=\bar{p}^{+}\left(X\mp\xi\right)\left(u_{a}^{1,2}-u_{b}^{1,2}%
\right)  \label{app4}
\end{equation}
\begin{equation}
\rho_{1,2}=\bar{p}^{+}\left(X\mp1\right)\left(u_{a}^{1,2}-u_{c}^{1,2}\right)
\label{app5}
\end{equation}

\item $\xi\leq X\leq1$

\begin{equation}
\epsilon_{1}=\frac{1}{X-1},\quad\epsilon_{2}=0
\end{equation}
\begin{equation}
\alpha_{1}=\frac{X-\xi}{X-1},\quad\gamma_{1}=\frac{X+\xi}{X-1}
\end{equation}
\begin{equation}
\beta_{1}=\bar{p}^{+}\left(X-\xi\right)\left(u_{c}^{1}-u_{b}^{1}\right)
\end{equation}
\begin{equation}
\rho_{1}=\bar{p}^{+}\left(X+\xi\right)\left(u_{c}^{1}-u_{a}^{1}\right)
\label{app6}
\end{equation}
\end{itemize}

Powers of $z_{i}$-s in the numerator can also be different in each interval.
They read

\begin{itemize}
\item $-1\leq X<-\xi$ and $\xi\leq X\leq1$

\begin{equation}
p_{1,2}=3n,\qquad r_{1,2}=4n,\qquad s_{1,2}=n
\end{equation}
\begin{equation}
a_{1,2}=3n,\qquad b_{1,2}=2n,\qquad c_{1,2}=3n
\end{equation}
\begin{equation}
d_{1,2}=n,\qquad e_{1,2}=4n,\qquad g_{1,2}=3n
\end{equation}

\item $-\xi\leq X<\xi$

\begin{equation}
p_{1}=n,\qquad r_{1}=3n,\qquad s_{1}=4n   \label{app7}
\end{equation}
\begin{equation}
a_{1}=3n,\qquad b_{1}=2n,\qquad c_{1}=3n
\end{equation}
\begin{equation}
d_{1}=3n,\qquad e_{1}=n,\qquad g_{1}=4n
\end{equation}
\begin{equation}
p_{2}=n,\qquad r_{2}=4n,\qquad s_{2}=3n
\end{equation}
\begin{equation}
a_{2}=3n,\qquad b_{2}=2n,\qquad c_{2}=3n
\end{equation}
\begin{equation}
d_{2}=3n,\qquad e_{2}=4n,\qquad g_{2}=n   \label{app8}
\end{equation}
\end{itemize}

For negative $\xi$ we have to change from Eqs. (\ref{app1}-\ref{app6}) only
Eqs. (\ref{app2}-\ref{app5}). Appropriate expressions read%
\begin{equation}
\epsilon_{1,2}=\mp\frac{1}{X\mp\xi}
\end{equation}
\begin{equation}
\alpha_{1,2}=\frac{X\pm\xi}{X\mp\xi},\quad\gamma_{1,2}=\frac{X\mp1}{X\mp\xi}
\end{equation}

\begin{equation}
\beta_{1,2}=\bar{p}^{+}\left( X\pm\xi\right) \left(
u_{b}^{1,2}-u_{a}^{1,2}\right)
\end{equation}
\begin{equation}
\rho_{1,2}=\bar{p}^{+}\left( X\mp1\right) \left(
u_{b}^{1,2}-u_{c}^{1,2}\right) .
\end{equation}
Also Eqs. (\ref{app7}-\ref{app8}) have to be replaced by%
\begin{equation}
p_{1,2}=4n,\qquad r_{1,2}=n,\qquad s_{1,2}=3n
\end{equation}
\begin{equation}
a_{1,2}=2n,\qquad b_{1,2}=3n,\qquad c_{1,2}=3n
\end{equation}
\begin{equation}
d_{1,2}=4n,\qquad e_{1,2}=3n,\qquad g_{1,2}=n
\end{equation}

In the case of ATDA the general formula for $\mathcal{J}_{1,2}$ is the same
as for VTDA, but now we have different expressions for $A_{1,2},\,
B_{1,2},\, C_{1,2}$. They are%
\begin{equation}
A_{1,2}=g^{\mp}\pm2v^{\mp}
\end{equation}
\begin{equation}
B_{1,2}=-f^{\mp}-g^{\mp}
\end{equation}
\begin{equation}
C_{1,2}=-2g^{\mp}\mp2v^{\mp}+f^{\mp}\pm1,
\end{equation}
where%
\begin{equation}
f^{\pm}=\pm1+X-\frac{2\xi\Lambda^{2}\vec{\kappa}_{T}\cdot\vec{q}_{T}}{\left(
1-\xi^{2}\right) t}
\end{equation}
\begin{equation}
g^{\pm}=-\frac{1}{2}\left( X\pm1\right) -\frac{\Lambda^{2}\vec{\kappa}%
_{T}\cdot\vec{q}_{T}}{\left( 1+\xi\right) t}
\end{equation}
\begin{multline}
v^{\pm}=\frac{1}{2\left( 1-\xi\right) \left( \xi+1\right) ^{2}t^{2}}\bigg\{%
8\xi\left( \Lambda\vec{\kappa}_{T}\cdot\vec{q}_{T}\right) ^{2} \\
+2\Lambda\left( 2\xi^{2}+\xi-1\right) t\left( X\pm1\right) \vec{\kappa }%
_{T}\cdot\vec{q}_{T} \\
+\left( \xi^{2}-1\right) t\left( \left( \xi+1\right) \left( X\pm1\right)
^{2}t-4\Lambda\xi\kappa_{T}^{2}\right) \bigg\}.
\end{multline}
Powers of $z_{i}$-s are the same as in the vector case.


\begin{thebibliography}{99}
\bibitem{PirSzym} B. Pire, L. Szymanowski, Phys. Rev. \textbf{D} 71 (2005)
111501, [ {hep-ph/0411387}]; J.~P.~Lansberg, B.~Pire and
L.~Szymanowski,
Phys.\ Rev.\ \textbf{D} 73 074014 (2006), [ {hep-ph/0602195}].

\bibitem{LanPirSzy} J.P. Lansberg, B. Pire, L. Szymanowski, %
 {0709.2567~[hep-ph]}.

\bibitem{Tiburzi} B.C. Tiburzi, Phys. Rev. \textbf{D} 72 (2005) 094001, [%
 {hep-ph/0508112}].

\bibitem{Courtoy} A. Courtoy, S. Noguera, Phys.\ Rev.\ \textbf{D} 76 (2007)
094026, [ {0707.3366~[hep-ph]}].

\bibitem{RAB} E. Ruiz Arriola, W. Broniowski, Phys. Lett. \textbf{B} 649
(2007) 49, [ {hep-ph/0701243}].

\bibitem{MPAR1} M. Praszalowicz, A. Rostworowski, Phys. Rev. \textbf{D} 64
(2001) 074003, [ {hep-ph/0105188}]; Phys. Rev. \textbf{D} 66 (2002)
054002, [ {hep-ph/0111196}].

\bibitem{Kimlocal} S.-i.~Nam, H.-C.~Kim, A.~Hosaka, M.M.~Musakhanov,
Phys.\ Rev.\ \textbf{D} 74 (2006) 014019, [ {hep-ph/0605259}];
S.~i.~Nam and H.~C.~Kim,
Phys.\ Rev.\ \textbf{D} 74 (2006) 096007, [ {hep-ph/0608018}].

\bibitem{MPAR2} M. Praszalowicz, A. Rostworowski,
Acta Phys.Polon. \textbf{B} 34 (2003) 2699, [ {hep-ph/0302269}].

\bibitem{Bochum} V.Yu. Petrov and P.V. Pobylitsa,  {hep-ph/9712203}.

\bibitem{Bochum1} V.Yu. Petrov, M.V. Polyakov, R. Ruskov, Ch. Weiss, K.
Goeke, Phys. Rev. \textbf{D} 59 (1999) 114018, [ {hep-ph/9807229}].

\bibitem{Christov}  C.~V.~Christov \textit{et al.},
Prog.\ Part.\ Nucl.\ Phys.\ \textbf{37} (1996) 91  [ {hep-ph/9604441}].

\bibitem{RipBroGo}  W.~Broniowski, B.~Golli and G.~Ripka,
Nucl.\ Phys.\ \textbf{A} 703 (2002) 667  [ {hep-ph/0107139}];
Phys.\ Lett.\ \textbf{B} 437 (1998) 24  [arXiv:hep-ph/9807261].

\bibitem{Blotzetal}  A.~Blotz, D.~Diakonov, K.~Goeke, N.~W.~Park, V.~Petrov
and P.V.~Pobylitsa,
Nucl.\ Phys.\ \textbf{A} {\ 555} (1993) 765.

\bibitem{Radyushkin:1999ms} A.~V.~Radyushkin,
Acta Phys.\ Pol.\ \textbf{B} 30 (1999) 3647, [ {hep-ph/0011383}].

\bibitem{Bochum2pi} M.V. Polyakov, Ch. Weiss,
Phys.Rev. \textbf{D} 59 (1999) 091502, [ {hep-ph/9806390}]; Phys.Rev.
\textbf{D} 60 (1999) 114017, [ {hep-ph/9902451}].

\bibitem{SQM} E. Ruiz Arriola, W. Broniowski, Phys.Rev. \textbf{D} 67 (2003)
074021, [ {hep-ph/0301202}].

\bibitem{Davidson} R.M.~Davidson, E.~Ruiz Arriola,
Acta Phys.\ Pol.\ \textbf{B} 33 1791 (2002), [ {hep-ph/0110291}].

\bibitem{PagelsStokar}  H.~Pagels and S.~Stokar,
Phys.\ Rev.\ D \textbf{20} (1979) 2947.

\bibitem{BallChiu}  J.~S.~Ball and T.~W.~Chiu,
Phys.\ Rev.\ \textbf{D} {\ 22} (1980) 2542.

\bibitem{Holdom}  B.~Holdom, J.~Terning and K.~Verbeek,
Phys.\ Lett.\ \textbf{B} {\ 232} (1989) 351;  Phys.\ Lett.\ \textbf{B} {\ 245%
} (1990) 612.

\bibitem{Birse} R.D. Bowler, M. Birse, Nucl. Phys. \textbf{A} 582 (1995)
655, [ {hep-ph/9407336}]; R.S. Plant, M. Birse, Nucl. Phys. \textbf{A}
628 (1998) 607, [ {hep-ph/9705372}].

\bibitem{Frank}  M.~R.~Frank, K.L.~Mitchell, C.D.~Roberts and P.C.~Tandy,
Phys.\ Lett.\ \textbf{B} {\ 359} (1995) 17  [ {hep-ph/9412219}].

\bibitem{WBnl} W. Broniowski, in: Miniworkshop on Hadrons as Solitons, Bled
1999,  {hep-ph/9909438}.

\bibitem{dor} A.E. Dorokhov, L. Tomio, Phys.Rev. \textbf{D} 62 (2000)
014016; I.V. Anikin, A.E. Dorokhov, L.Tomio, Phys. Lett. \textbf{B} 475
(2000) 361, [ {hep-ph/9909368}].

\bibitem{KimNL} S.-i.~Nam and H.-C.~Kim,
Phys.\ Rev.\ \textbf{D} 74 (2006) 076005, [ {hep-ph/0609267}].


\bibitem{Noguera} S.~Noguera, 
Int.\ J.\ Mod.\ Phys.\ \textbf{E} 16 (2007) 97, [ {hep-ph/0502171}];
S.~Noguera and V.~Vento,
Eur.\ Phys.\ J.\ \textbf{A} 28 (2006) 227, [ {hep-ph/0505102}].


\bibitem{BzdakMP} A. Bzdak, M. Praszalowicz, Acta Phys. Polon. \textbf{B} 34
(2003) 3401, [ {hep-ph/0305217}].

\bibitem{Kimanomaly} M.M.~Musakhanov, H.-C.~Kim,
Phys.\ Lett.\ \textbf{B } 572 (2003) 181, [ {hep-ph/0206233}].

\bibitem{DP} D.I. Diakonov, V.Yu. Petrov, Nucl. Phys. \textbf{B} 245 (1984)
259; \textbf{B} 272 (1986) 457.

\bibitem{DPrev} D.I. Diakonov, V.Yu. Petrov, [ {hep-ph/0009006}].

\bibitem{gradient} M.~Praszalowicz and G.~Valencia,
Nucl.\ Phys.\ \textbf{B} 341(1990) 27; 
E.~Ruiz Arriola,
Phys.\ Lett.\ \textbf{B} 253 (1991) 430.

\bibitem{Ametller}  L.~Ametller, L.~Bergstrom, A.~Bramon and E.~Masso,
Nucl.\ Phys.\ \textbf{B} {\ 228} (1983) 301.

\bibitem{Gronberg} J. Gronberg et al. (CLEO), Phys. Rev. D57, 33 (1998), [%
 {hep-ex/9707031}].

\bibitem{Courtoy1}  A.~Courtoy and S.~Noguera,
 {arXiv:0804.4337 [hep-ph]}.  

\bibitem{BAK} W.~Broniowski, E.~R.~Arriola and K.~Golec-Biernat,
Phys.\ Rev.\ \textbf{D} 77 (2008) 034023, [ {0712.1012 [hep-ph]\,]}.
\end{thebibliography}
\end{document}